\documentclass[article,11pt]{emulateapj}
\usepackage{lineno}
\usepackage{graphicx}
\usepackage{times}
\usepackage{natbib}
\usepackage{amsmath}
\usepackage{amssymb}

\def\jgr{J. Geophys. Res. }

\def\aj{Astron. J.}
\def\asht{AsH$_3$ }
\def\ashtx{AsH$_3$}
\def\apj{Astrophys. J.}
\def\apjl{Astrophys. J. Lett.}

\def\jqsrt{J. Quant. Spectrosc. \& Rad. Transf. }

\def\icm{cm$^{-1}$ }

\def\deg{$^\circ$ }
\def\degx{$^\circ$}

\def\mum{$\mu$m }
\def\mumx{$\mu$m}
\def\chf{CH$_4$ }
\def\chfx{CH$_4$}
\def\chtd{CH$_3$D }
\def\chtdx{CH$_3$D}
\def\chisq{$\chi^2$ }
\def\chisqx{$\chi^2$}
\def\herat{He/H$_2$ }
\def\heratx{He/H$_2$}
\def\pht{PH$_3$ }
\def\phtx{PH$_3$}

\def\tenmsix{$\times 10^{-6}$ }
\def\tenmsixx{$\times 10^{-6}$}
\def\nht{NH$_3$ }
\def\nhtx{NH$_3$}

\begin{document}
\title{Cloud clearing in the wake of Saturn's Great Storm of 2010 -- 2011 and
suggested new constraints on Saturn's He/H$_2$ ratio.}
\author{L.A. Sromovsky$^1$, K. H. Baines$^1$, P.M. Fry$^1$, and T. W. Momary$^2$}
\affil{$^1$Space Science and Engineering Center, University of Wisconsin-Madison,
1225 West Dayton Street, Madison, WI 53706, USA}
\affil{$^2$Jet Propulsion Laboratory, 4800 Oak Grove Drive, Pasadena, CA 91109, USA}


\slugcomment{Journal reference: Icarus 276 (2016) 141-142}
\begin{abstract}
Saturn's Great Storm of 2010 -- 2011 produced a planet-encircling wake
that slowly transitioned from a region that was mainly dark at 5 \mum
in February 2011 to a region that was almost entirely bright and
remarkably uniform by December of 2012. The uniformity and high
emission levels suggested that the entire wake region had been cleared
not only of the ammonia clouds that the storm had generated and
exposed, but also of any other aerosols that might provide significant
blocking of the thermal emission from Saturn's deeper and warmer
atmospheric layers.  Our analysis of VIMS wake spectra from December
2012 provides no evidence of ammonia ice absorption, but shows that at
least one significant cloud layer remained behind: a non-absorbing
layer of 3 -- 4 optical depths (at 2 \mumx) extending from 150 to
$\sim$400 mbar.  A second layer of absorbing and scattering particles,
with less than 1 optical depth and located near 1 bar, is also
suggested, but its existence as a model requirement depends on what
value of the \herat ratio is assumed. The observations can be fit well
with just a single (upper) cloud layer for a \herat ratio
$\approx$0.064 in combination with a \pht deep volume mixing ratio of
5 ppm.  At lower \herat ratios, the observed spectra can be modeled
without particles in this region. At higher ratios, in order to fit
the brightest wake spectrum, models must include either significant
cloud opacity in this region, or significantly increased absorption by
\phtx, \nhtx, and \ashtx.  As the exceptional horizontal uniformity in
the late wake is most easily understood as a complete removal of a
deep cloud layer, and after considering independent constraints on
trace gas mixing ratios, we conclude that the existence of this
remarkable wake uniformity is most consistent with a He/H$_2$ mixing
ratio of 0.055$^{+0.010}_{-0.015}$, which is on the low side of the
0.038 -- 0.135 range of previous estimates.

\end{abstract}

\maketitle
\shortauthors{Sromovsky et al.} 
\shorttitle{Cloud clearing following Saturn's Great Storm.}
\newpage

\section{Introduction}

Saturn's Great Storm of 2010 -- 2011 was one of the most powerful
convective events ever witnessed.  It's rapid development, its huge
horizontal scale, and the planet encircling wake it generated within
$\sim$6 months were accompanied by a prolific generation of lightning
\citep{Fischer2011,Dyudina2013} and an unusual spectral character
consistent with the delivery of ammonia and water ices to the visible
cloud deck \citep{Sro2013gws}, which is $\sim$200 km above the water
vapor condensation level near 20 bar, where the lightning also appears
to have originated.  The dramatic growth of the storm after its
initial formation in early December 2010 was documented by amateur and
professional groundbased imaging \citep{Sanchez-Lavega2012} and by
Cassini imaging with the Imaging Science Subsystem (ISS)
\citep{Sayanagi2013}.

\begin{figure*}[!hbt]\centering
\includegraphics[width=6in]{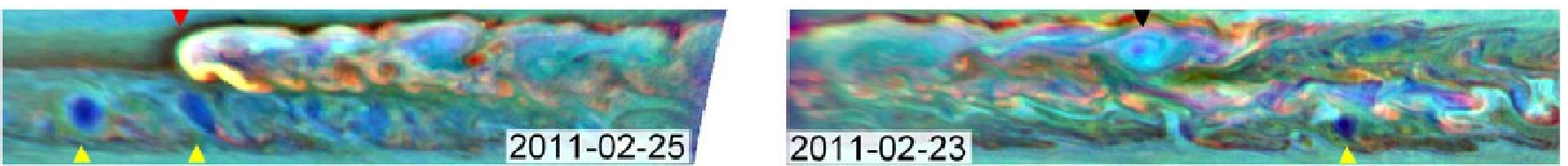}
\includegraphics[width=6in]{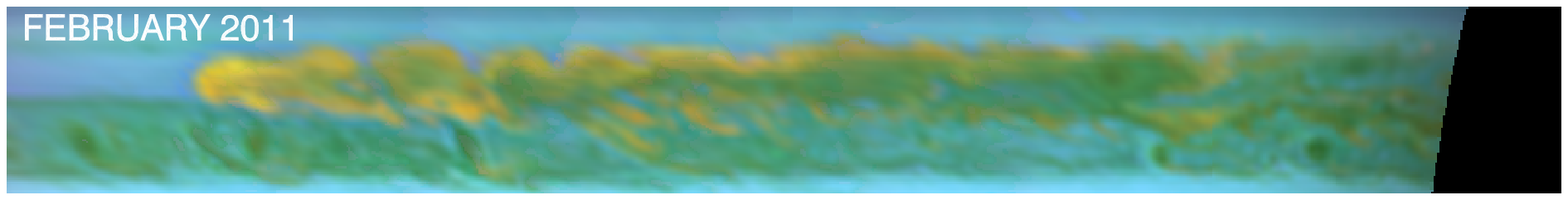}
\includegraphics[width=6in]{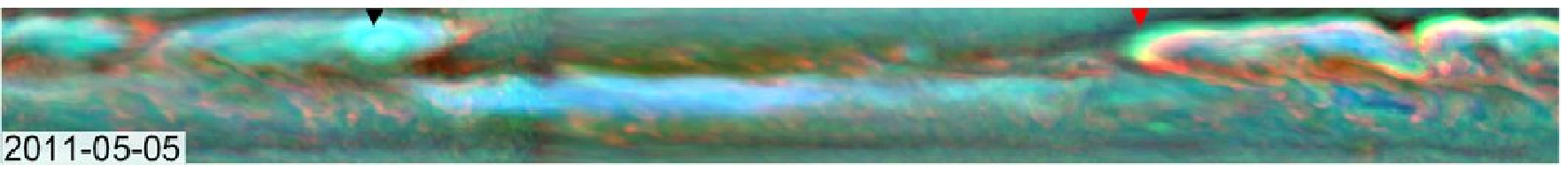}
\includegraphics[width=6in]{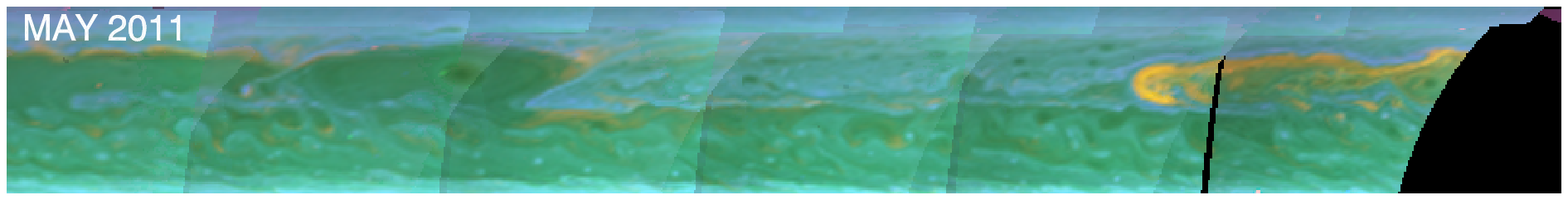}
\caption{Early stages of the Great Storm in February 2011 (top 2
  panels). The first and third panels from the top, from
  \cite{Sayanagi2013}, are mosaics of ISS observations displayed using
  CB2, MT2, and MT3 filters for R, G, and B channels, in which red
  triangles indicate the location of the storm head, black triangles
  the location of the large bright blue anticyclonic oval vortex, which
Sayanagi et al. referred to as the AV, and yellow
  triangles the locations of dark ovals. The storm head disappeared
  after it caught up to the anticyclone in June 2011.  The second and
  fourth panels are from VIMS, covering 200\deg of longitude and
  latitudes 20\deg N to 44 \deg N. The VIMS color maps used R = 4.08
  \mumx, G = 1.89 \mumx, and B = 3.05 \mumx, for which yellow/orange
  regions indicate large particles, optically thick clouds, and strong
  3-\mum absorption characteristic of ammonia ice.
Note that features may appear at slightly different longitudes in 
pairs of maps from the same month due to time differences.
\label{Fig:moscomp}}
\end{figure*}

The morphological characteristics of the storm during February and May
of 2011 are illustrated in Fig.\ \ref{Fig:moscomp}, where ISS images
are displayed in
comparison with infrared images obtained by the Visual and Infrared
Mapping Spectrometer (VIMS).  The first VIMS spectral imaging of the
storm occurred in February 2011, after it was well developed and had a
planet-encircling wake, though at the latitude of the storm head, the
region upstream of the head was still undisturbed at that time. In
these VIMS images, color assignments are 4.08 \mum for red, 1.89 \mum
for green, and 3.05 \mum for blue.  The strong absorption at 3 \mum in
clouds that are bright at the other two wavelengths produce the orange
color that indicates the presence of ammonia ice, which was 
mainly confined to a 10\deg band of latitude centered at
35\degx N planetocentric latitude, and much less evident in the secondary
wake extending south of the storm, especially in the May 2011 images.

\begin{figure*}[!hbt]\centering
\includegraphics[width=6in]{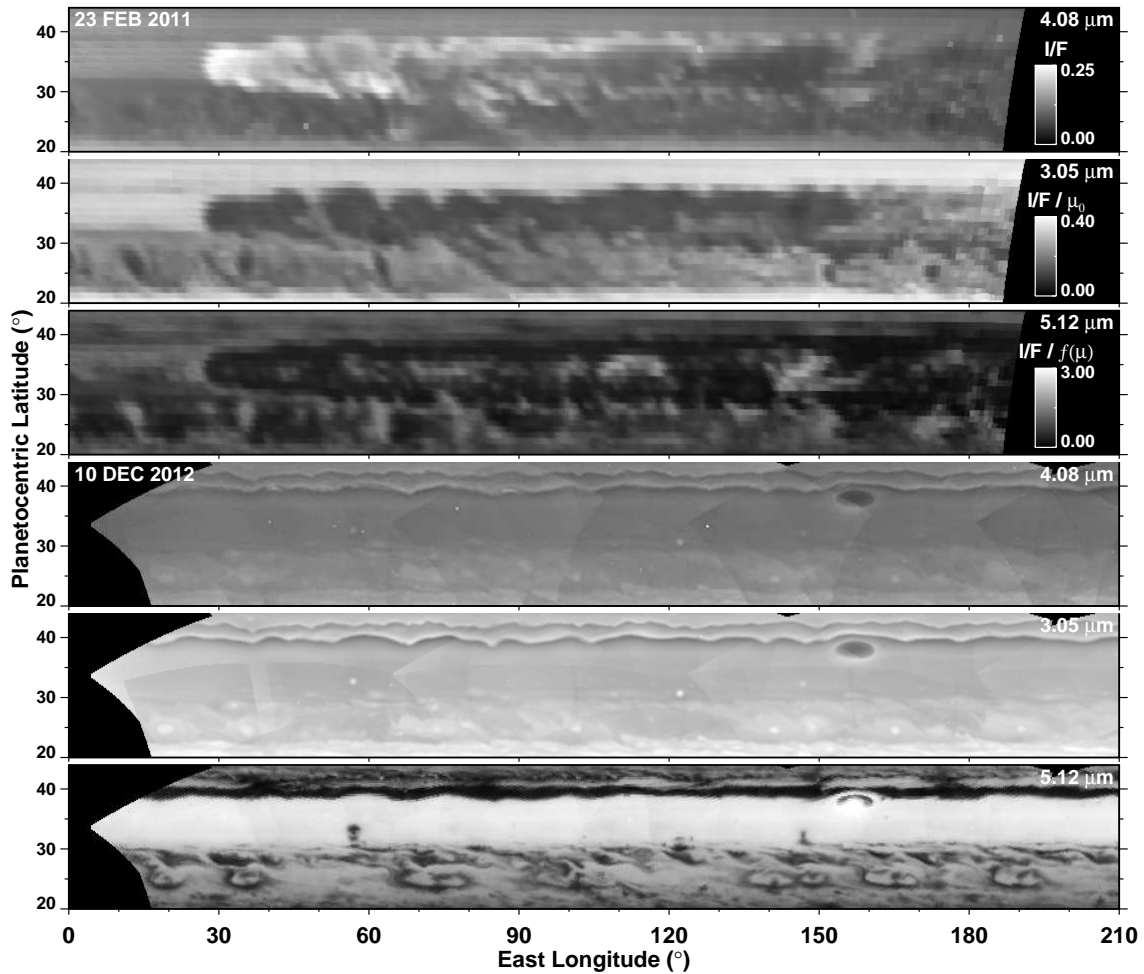}
\caption{Comparison of storm latitudes at 3.05, 4.08, and 5.12 \mumx,
  on 23 February 2011 (top three panels) and 10 December 2012 (bottom
  three panels). In the 23 February 2011 images, we see that clouds
  with large particles, indicated by features that are bright at 4.08
  \mumx, are also clouds with a significant ammonia ice component,
  indicated by absorption at 3.05 \mumx, and are optically thick
  enough (and absorbing enough) to block thermal emission at 5
  \mumx. On the other hand, in the 10 December 2012 images there is no
  evidence of large particles, no evidence of ammonia absorption, and
  a surprisingly uniform lack of blocking of 5-\mum thermal
  emission. The gray scales used for the bottom three panels are the
  same as for the corresponding top three panels. Mosaics were blended
  with approximate limb darkening corrections: at 3.05 \mum I/F values
  were divided by the cosine of the solar zenith angle; at 5.12 \mum
  I/F values were divided by $f(\mu) = 0.427-0.445\mu+1.019\mu^2$, where
  $\mu$ is the cosine of the observer zenith angle; no correction was
  made at 4.08 \mumx.\label{Fig:345}}
\end{figure*}

The sample VIMS data set provided in Fig.\ \ref{Fig:345} emphasizes
the most striking spectral characteristic of the storm: its remarkably
low reflectivity at wavelengths near 3 \mumx. (Here brightness is
expressed in units of I/F, given by radiance/(solar irradiance/$\pi$),
which can exceed unity near 5 \mum because there Saturn's thermal
emission can exceed the amount of reflected sunlight.)  Clearly, the
materials convected upward from deeper levels are very different from
the surrounding clouds that usually dominate Saturn, which have no
trace of 3-\mum absorption.  From a spectral analysis of the storm
head \cite{Sro2013gws} showed that it contained a mixture of primarily
ammonia ice, with likely contributions of water ice, and a third
component that might be either NH$_4$SH or the unknown material that
dominates the upper haze over most of Saturn. As evident in
Fig.\ \ref{Fig:moscomp}, the ammonia absorption was widespread in the
main wake region when the storm head was in a highly active state. We
can also see that at 4.08 \mumx, a pseudo continuum wavelength, the
storm clouds are brighter than surrounding clouds, indicating
relatively larger particles. Where they are bright at 4 \mumx, they
are also optically thick enough and absorbing enough at 5 \mum to
block thermal emission from deeper layers of Saturn's atmosphere.  The
wake region in February 2011 also had regions where 5-\mum emission
was greater than regions upstream of the storm, one of which (near
150\deg east) is near the long-lived anticyclonic vortex (AV)
described by \cite{Sayanagi2013} and \cite{Momary2014} and may be an
effect produced by that circulation feature.

Figure \ref{Fig:345} also shows that a remarkable transition of the
wake region occurred during the the two years following its beginning
in early December 2010.  By December 2012, in the latitude band from
30\deg N to 39\deg N there was no evidence of bright large-particle
cloud features at 4.08 \mum and no evidence of 3-\mum absorption by
ammonia ice (or anything else). Most surprising, and a unique outer
planet feature as far as we know, the primary wake region turned from
being mainly very dark at 5 \mum to being entirely bright and
remarkably uniform.   This indicates a dramatic
decrease in the opacity of aerosols that normally attenuate thermal
radiation emanating from the 5 -- 6 bar level in Saturn's atmosphere.
I/F values at 5 \mum increased from pre-storm values of 0.6 -- 0.8,
and even lower deep storm values of 0.2 -- 0.3, to a remarkably high
I/F averaging $\sim$2.7.   The spatial uniformity and high emission
levels reached by December 2012 suggested that the entire wake region
had been cleared not only of the ammonia clouds that the storm had
generated, but also of any other deep aerosols that might provide
significant blocking of Saturn's 5-\mum thermal emission.  

In the following, we use VIMS imaging at 5 \mum to define the
morphological evolution of the wake's ``cleared'' regions.  We then
use VIMS spectral observations to constrain the cloud structure of
those regions, finding that they were not completely cleared of all
aerosols, but instead retained an upper level cloud similar to
surrounding regions. We show that a deep layer of aerosols that blocks
part of the thermal emission declined dramatically to less than one
optical depth, and is only needed if the He/H$_2$ mixing ratio is at
the higher end of the range of values previously published, but find that
complete deep clearing could explain the remarkable uniformity of the
late wake region if the He/H$_2$ mixing ratio is in the lower part of
that range.
 
\section{Overview of wake evolution}

\subsection{Evolution of apparent wake ``clearing''}

\begin{figure*}[!hbt]\centering
\includegraphics[width=6.25in]{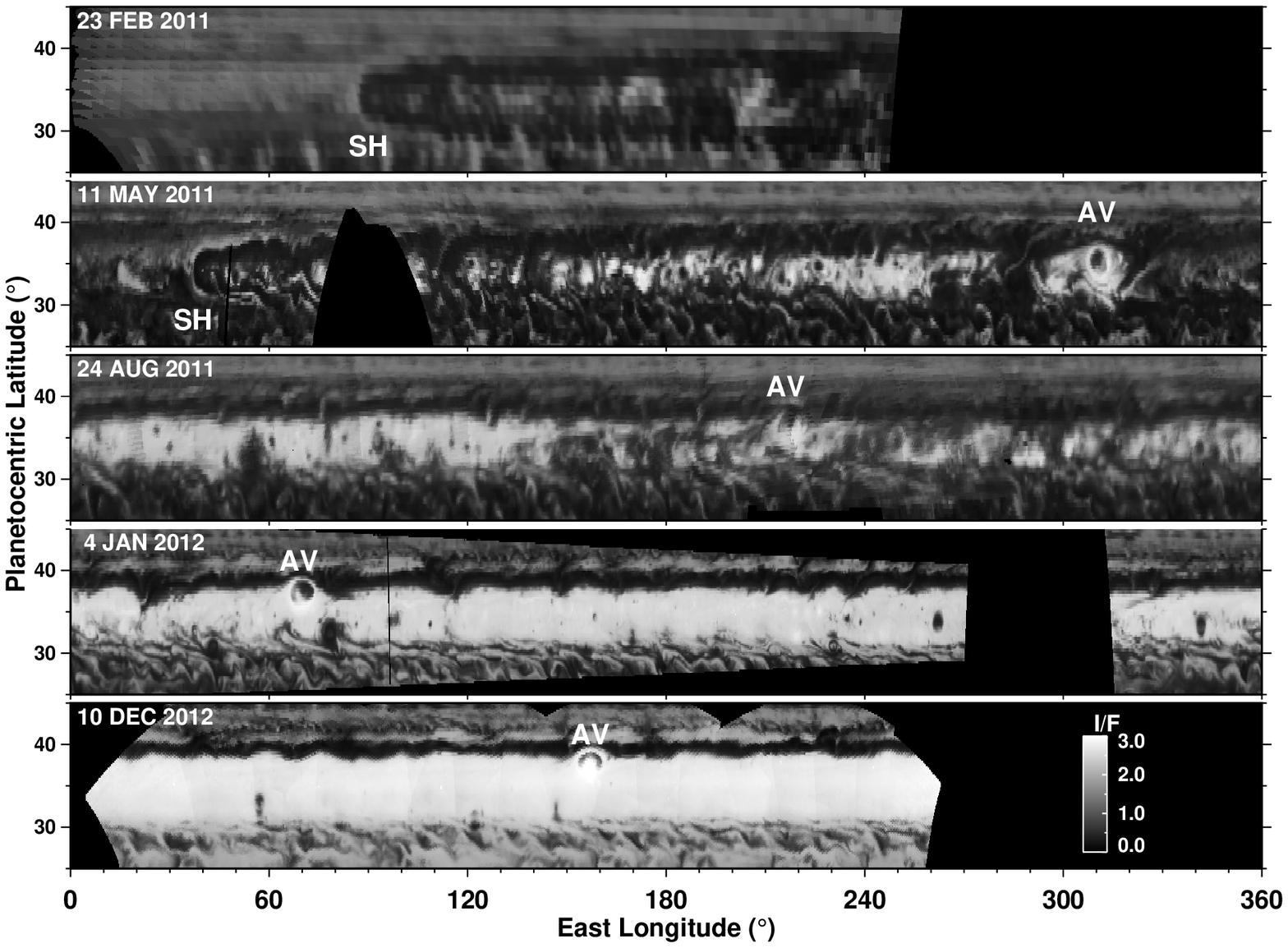}
\caption{Rectilinear 5-\mum maps showing the evolution of apparent
  wake-clearing from 23 February 2011 to 10 December 2012.  All images
  are shown with the same gray scale indicated for the December 2012
  image. For easier comparison with wake structure in May, the
  February map was displayed 160\deg west of its actual position. In
  four of the images we marked the position of the anticyclonic vortex
  (AV) that developed along with the Great Storm. The storm head (SH)
  is marked in the top two images, but not in subsequent images
  because it dissipated after encountering the AV in mid
  June.\label{Fig:5mos}}
\end{figure*}

The morphological evolution of the wake is illustrated by the 5-\mum
mosaics displayed in Fig.\ \ref{Fig:5mos}.  After the anticyclone was
overtaken by the head of the storm in mid June 2011
\citep{Sayanagi2013}, no evidence of the storm head was subsequently
seen.  But even before that event, the wake was already beginning to
develop regions of high emission at 5 \mumx, evident from the 11 May
2011 mosaic in Fig.\ \ref{Fig:5mos}.  It may also be significant that
the region around the anticyclone (near 320\degx E in that mosaic) is
marked by excess emission.  The widespread clearing seemed to begin in
local regions distributed near the mid line of the storm's main wake,
and over time became more widely distributed within the wake.  By
August 2011, the regions of excess 5-\mum emission grew significantly
in number and both in longitudinal coverage and in latitudinal extent.
By December 2012 the 5-\mum bright wake region spanned latitudes from
30\deg N to 39\deg N (planetocentric), extended over all longitudes,
and became nearly uniform in brightness, suggesting the possibility of
complete removal of the cloud layer that was strongly attenuating
thermal emissions from the deeper layers of Saturn.  However, we will
show in what follows that the cloud clearing explanation is only
plausible for a limited range of He/H$_2$ mixing ratios.
  
\subsection{Development of spatial uniformity in the cleared wake region}

A quantitative measure of the developing spatial uniformity of the wake region 
is provided by longitudinal
scans at 35\deg N for each of the mosaics displayed in
Fig.\ \ref{Fig:5mos}.  These scans show that I/F values at 5.12 \mum
(as corrected for view angle variations) were $\sim$1.0 upwind of the
storm head in February 2011, but ranged from 0.3 to
1.9 downstream of the storm head, with relatively small variability
ahead of the storm and much larger variability in the wake region. The
wake variability at this wavelength increased dramatically by May 2011,
both in amplitude of variation (ranging from $\sim$0.2 to $\sim$2.7)
and in spatial frequency.  But by January 2012, as the wake extended
over most longitudes, the I/F in the wake region grew much more
uniform, with most values between 2.4 and 2.6. Finally, by December of
2012, the mean wake I/F reached 2.75 with a standard deviation of only
2\%.  A plausible qualitative explanation of this behavior is that the
tropospheric temperature structure remained relatively stable both
longitudinally and over time, but that the deep ammonia cloud layer
present before the storm became disturbed by heterogeneous convective
towers during the storm, leading to reduced emission where clouds were
developing and increased emission where local downwelling motions
produced a reduction in deep cloud optical depth.  This perhaps led to
strong variability while the storm clouds were active.  But after the
storm subsided, the deep clouds were completely cleared, perhaps by
the mechanism of \cite{Li2015}.  Whether this interpretation is
quantitatively consistent with cloud structure inferred by radiative
transfer modeling is treated in a subsequent section. The fact
that the peak I/F seen in May 2012 is essentially the same as the mean
seen in the December 2012 mosaic suggests that the main difference
between them is the fraction of cloud clearing rather than changes
in thermal structure.

\section{Overview of VIMS spectral observations}

\subsection{Observations list}

Our work is based on VIMS observations gathered from February 2011
(the first VIMS observations of the storm region) through December
2012 (when the cleared region reached full longitudinal and
latitudinal extent). Data sets that were analyzed and observing
conditions are listed in Table\ \ref{Tbl:obslist}. These all include
measurements of reflected sunlight as well as thermal emission,
which we found to provide the best combination of constraints on upper and lower cloud
parameters and \phtx.

\begin{table*}\centering
\caption{List of RC17-calibrated VIMS data cubes and observing conditions used in our analysis.}
\vspace{0.1in}
\begin{tabular}{ r c c c r c}
                             &                   &     UT Date  &        Start &   Pixel &    Phase \\
              Observation ID &     Cube Version &   {\footnotesize YYYY-MM-DD} &      Time &     size & angle \\
\hline
       VIMS\_145SA\_WIND5HR001 &   V1677201862\_3 &   2011-02-24 &     00:36:35 &     882 km &    51\deg \\
 VIMS\_148SA\_NHEMMAP001\_PRIME &   V1683829310\_1 &   2011-05-11 &     17:33:22 &    331 km&    23\deg \\
VIMS\_152SA\_PEARLMOV001\_PRIME &   V1692862427\_1 &   2011-08-24 &     06:44:21 &    403 km&    15\deg \\
VIMS\_159SA\_HIRESMAP001\_PRIME &   V1704401716\_1 &   2012-01-04 &     20:04:37 &    157 km&    56\deg \\
        VIMS\_160SA\_MIRMAP001 &   V1705983395\_1 &   2012-01-23 &     03:25:46 &    1137 km&    74\deg \\
 VIMS\_176SA\_NORSTRM001\_PRIME &   V1733862483\_1 &   2012-10-12 &     19:34:16 &    274 km&    36\deg \\
 VIMS\_176SA\_NORSTRM001\_PRIME &   V1733869346\_1 &   2012-12-10 &     21:28:39 &    288 km&    31\deg \\
\hline
\end{tabular}\label{Tbl:obslist}
\end{table*}

\subsection{VIMS instrumental characteristics}

 VIMS is a 2-channel mapping spectrometer, one channel covering the
 0.35 -- 1.0 \mum spectral range, and the second covering an overlapping near-IR
 range from 0.85 -- 5.1 \mumx. The near-IR spectral range is covered by 256
 contiguous wavelength bands sampled at a nominal interval of 0.016 \mumx.  Each
 spatial sample (pixel) has a square field of view 0.5 milliradians on
 a side, and a typical image frame has dimensions of 64 pixels by 64
 pixels. Detailed descriptions of the VIMS instrument are provided by
 \cite{Brown2004} and \cite{McCord2004}.  All wavelengths are measured
 simultaneously.  Spatial samples are obtained via raster scans of the
 instrument field of view across the target.  Random noise is a very
 small contributor to uncertainty in the VIMS observations. It is
 systematic effects that are the main source of uncertainties in
 deriving atmospheric constraints from the VIMS spectra. Many are
 related to calibration uncertainties.

\subsection{VIMS calibration}

There are four aspects of the VIMS calibration that merit special
attention.  First, is the wavelength scale and its variation over
time.  \cite{Sro2013gws} noted that the wavelengths of the VIMS
spectral channels had changed typically by 5 -- 10 nm relative to the
assignments given in data headers.  Here we used a scaled version of
the shifts given by \cite{Sro2013gws}.  These were based on alignment
of atmospheric gas absorption features with well known wavelengths.  A
second aspect is the correction for responsivity depressions in the
vicinity of the joints in order sorting filters on the VIMS detector
array, which occur near wavelengths of 1.69 \mumx, 2.98 \mumx, and
3.85 \mum \citep{Miller1996SPIE,Brown2004}.  The smallest effect on
responsivity is at the 2.98-\mum joint, which can be corrected for
following \cite{Sro2013gws}.  Effects at the other two joints are much
larger and not reliably correctable, and thus not used in constraining
model cloud structures.  A minor aspect is the correction of nominal
exposures for clock rate and time offset.  From comparison of images
of Jupiter at exposures of 20 ms, 80 ms, and 160 ms, we found that the
previous rule in which T(corrected) = T(nominal)$\times$1.0175 - 4 ms needed
to be changed to the T(corrected) = T(nominal)$\times$1.0175 - 1.67
ms. Without the change, I/F values computed for 60 ms exposures are
about 10\% too large.  As most exposures are much longer, this is not
generally an important correction.  The fourth issue has to do with
the radiometric calibration, which has been updated several times over
the years.  \cite{Clark2012} described a new radiometric calibration,
termed RC17, which was released to PDS and ISIS3 in mid 2014. That is
the radiometric calibration used in our current analysis.  Flat-field
correction files have also been revised over time.  In this analysis
we used the 2009 flat named ir\_flat\_3\_2009.cub in PDS volumes and
ir\_flatfield\_v0002.cub in ISIS3. The ISIS system is described by
\cite{Anderson2004}.

\section{Radiative transfer modeling}

Our radiative transfer calculations follow essentially the approach
used by \cite{Sro2013gws}, using the same multiple scattering code that
simultaneously accounts for thermal emission and reflected sunlight. A
revision of the code has enhanced its ability to use parallel
processing to speed calculations. While the prior code did each of the
ten correlated-k terms in parallel, the revision allows multiple
wavelengths to be run in parallel as well.  A second exception is that
we treated the thermal profile differently, as described in a
following section. A third exception is how we handled line-spread
functions, which is also described in a separate section.  Here we
summarize our assumed composition and gas absorption models.

\subsection{Atmospheric composition}

Measured by their effects on Saturn's 1 -- 5 \mum spectrum, the most
important minor gases are methane (\chf and \chtdx), phosphine (\phtx), arsine
(AsH$_3$), and ammonia (\nhtx).  In reflected sunlight
($\lambda <$ 4.6 \mumx), methane and phosphine are dominant, while in
thermal emission ($\lambda >$ 4.6 \mumx) phosphine is dominant, with
arsine and ammonia playing relatively minor roles. For \chf we used
the \cite{Fletcher2009ch4saturn} volume mixing ratio of
(4.7$\pm$0.2)$\times 10^{-3}$, which corresponds to a \chfx/H$_2$ ratio
of (5.3$\pm 0.2)\times 10^{-3}$.  For \chtd we also used the
\cite{Fletcher2009ch4saturn} VMR value of 3$\times 10^{-7}$.  The most
important variable gas is PH$_3$, and its vertical profile needs to be
adjusted to fit VIMS spectra. 
We followed \cite{Fletcher2009ph3} in defining a pressure break point $P_0$, below which (in
altitude) the mixing ratio is a constant $\alpha_0$ and above which the mixing
ratio falls off with a constant gas to pressure scale height ratio $f$, so that
the mixing ratio as a function of pressure can be written as \begin{eqnarray}
 \alpha(P) = \alpha_0 (P/P_0)^{(1 - f)/f} \quad \mathrm{for} \quad P<P_0.\label{Eq:prof1}
\end{eqnarray}
For profiles with an additional break point at $P_{1} < P_0$ we can write
the mixing ratio above that point as \begin{eqnarray}
  \alpha(P)= \alpha_0 (P_1/P_0)^{(1 - f)/f}(P/P_1)^{(1 - f_1)/f_1} \quad 
   \mathrm{for} \quad P<P_1\label{Eq:prof2},
\end{eqnarray}
with Eq. \ref{Eq:prof1} still applying for $P_1<P<P_0$.  In most cases
we used profiles with a single break point with scale height ratios
near 0.5, somewhat greater than the CIRS-based values of
\cite{Fletcher2009ph3} and much greater than values near 0.2 typical
of the \cite{Fletcher2011vims} results at comparable latitudes. For
our initial calculations we used a break point at $P_0$ = 550 mb and
deep mixing ratios of 4 -- 5 ppm, which are comparable to the
CIRS-based values. We also assumed an initial \herat ratio of 0.0638
\citep{Hanel1981Sci}, a constant \asht volume mixing ratio of 6 ppb, and
the \nht profile given by \cite{Prinn1984}, which has a constant deep
mixing ratio of 2.05$\times 10^{-4}$, based on
\cite{Courtin1984}.

\subsection{Gas absorption models}

\begin{figure*}[!htb]\centering
\hspace{-0.1in}
\includegraphics[width=6.4in]{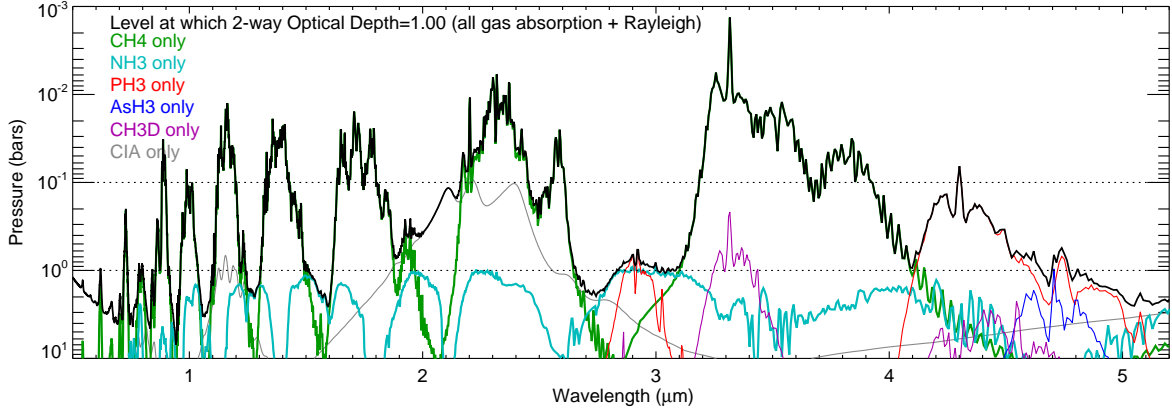}
\caption{Penetration depth of near-IR photons indicated by pressures
  at which a unit albedo reflecting layer produces an external I/F of
  1/e at normal incidence and viewing.  Curves are shown for methane
  only (green), ammonia only (cyan), phosphine only (red), arsine only
  (purple), CH$_3$D only (magenta), CIA only (gray), and all gases
  combined (black), assuming the \cite{Lindal1985} temperature
  profile, the ammonia profile of \cite{Prinn1984}, a \herat ratio of
  0.0638, a phosphine deep mixing ratio of 5 ppm, falling off above
  0.69 bars with a \pht to pressure scale height ratio of 0.5, an
  \asht VMR of 6 ppb, and a \chtd VMR of 0.3 ppm.}
\label{Fig:pendepth}
\end{figure*}

We limited our analysis to $\lambda>$1.268 \mum so that correlated-k
models for methane absorption could be based on line-by-line calculations, 
following \cite{Sro2012LBL}.
We used the same line data for
 computing \chtd absorption models.  For \nht we used the combined
 correlated-k absorption model described by \cite{Sro2010iso}, which
 is based primarily on the Goody-Lorentz band model of
 \cite{Bowles2008}.  Our exponential sum approximations of phosphine (\phtx) absorption 
 are based on the line data of
 \cite{Butler2006} in the 2.8 -- 3.1 \mum region and on the
 \cite{Rothman2009} HITRAN 2008 line data in the 4.1 -- 5.1 \mum
 region. Our model of AsH$_3$ absorption is based on line data from
 \cite{Tarrago1996} (via G. Bjoraker, via B. B\'ezard, personal
 communication).  Where multiple gases have overlapping absorptions
 we followed \cite{Lacis1991} to obtain 10-term
correlated-k approximations for the combined gases.
Collision-induced absorption
 (CIA) for H$_2$ and H$_2$-He was calculated using programs downloaded
 from the Atmospheres Node of the Planetary Data System, which are
 documented by \cite{Borysow1991h2h2f, Borysow1993errat} for the
 H$_2$-H$_2$ fundamental band, \cite{Zheng1995h2h2o1} for the first
 H$_2$-H$_2$ overtone band, and by \cite{Borysow1992h2he} for H$_2$-He
 bands.  
The diagnostic value of these gases can be estimated from their
penetration depths shown in Fig.\ \ref{Fig:pendepth}.
 Arsine has a noticeable effect on the VIMS
spectra at 4.5 -- 4.9 \mumx, which is where ammonia gas also plays a
relatively minor role.
Ammonia is more important in controlling thermal radiation at wavelengths
beyond 5.1 \mumx.  

\subsection{Thermal structure and the He/H$_2$ ratio}

\begin{figure}\centering
\includegraphics[width=3.5 in]{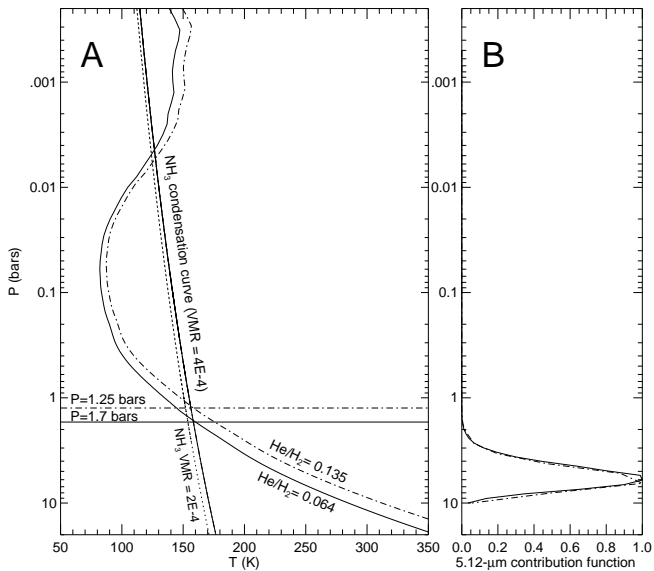}
\caption{A: Temperature profiles for two assumed He/H$_2$ ratios of
  0.064 (solid) and 0.135 (dot-dash). Also shown are the \nht
  saturation vapor pressure profile versus temperature for volume
  mixing ratios of 2$\times 10^{-4}$ and 4$\times 10^{-4}$ , and
  pressures at which \nht condensation can be expected to occur. B:
  Contribution functions $B_\lambda(T)\exp(-\tau)d\tau/d\ln P$
  normalized to the peak at 5.12 \mum for each temperature profile
  plotted in A. The peak contributions are from the 5.4 -- 5.9 bar
  level.}
\label{Fig:tprof}
\end{figure}

Voyager 2 radio occultation measurements \citep{Tyler1982Sci} were used to
determine refractivity profiles for Saturn.  From the refractivity
profiles, $T(P)$ profiles could be constructed for any given
assumed molecular composition.  The composition providing 
the best fits to IRIS spectral measurements in the 207 \icm -- 602 \icm 
spectral range, calculated over a range of latitudes, corresponded to a hydrogen
mole fraction of 0.940$\pm$0.005 \citep{Hanel1981Sci}, with an additional
absorption coefficient uncertainty of $\pm$0.005. Ignoring the minor effects of
heavier molecules, this leads to a He/H$_2$ ratio of 0.0638$\pm$0.007,
using the root sum of squares of the random and coefficient uncertainties.
Using this composition \cite{Lindal1985} derived a T(P) profile from
the Voyager 2 ingress refractivity profile down to about 800 mb, and
added trace amounts of NH$_3$ to constrain the profile down to 1.3 bars.
We extended this profile to 10 bars assuming an adiabatic lapse rate
of approximately -0.83 K/km.  To accommodate the possibility of other
He/H$_2$ ratios, we scaled this profile to preserve the same refractivity
profile, using the scaling relations \citep{Conrath2000}:
\begin{eqnarray}
T = T_0 \frac{m}{m_0}, \qquad P = P_0 \frac{m \alpha_0}{m_0 \alpha}, 
\end{eqnarray} 
where $T$, $P$, $m$, and $\alpha$ are the temperature, pressure,
molecular weight, and refractivity respectively of the modified
profile, and the same quantities with a zero subscript are for the
original profile.  Two sample profiles are shown in
Fig.\ \ref{Fig:tprof}A, which also shows how the different profiles
effect the pressure at which \nht would become saturated, using sample
\nht volume mixing ratios of 200 ppm and 400 ppm.  Note that the
condensation level moves to lower pressures for higher values of the
\herat ratio.  The reflection and emission spectra of Saturn are both
affected by the He/H$_2$ ratio because the collision-induced
absorption is dependent on the ratio, but the effect is most dramatic
on the thermal emission spectrum because of the strong dependence of
radiance on temperature in the 5-\mum region, where most of the emitted
radiation originates (Fig.\ \ref{Fig:tprof}B).  At low values of the
He/H$_2$ ratio the atmosphere becomes cool enough that the observed
5-\mum emission cannot be reproduced by even a cloud-free model.

\begin{figure}\centering
\includegraphics[width=3.5in]{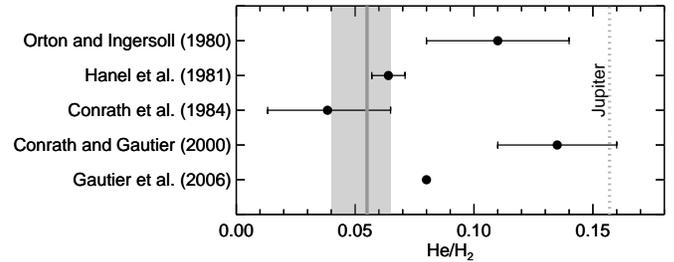}
\caption{Estimates of the \herat ratio on Saturn using methods described in the text. The value for
Jupiter (dotted line) is 0.157$\pm$0.003 \citep{VonZahn1998}. The range most
consistent with cloud clearing is shown by the gray band. \label{Fig:herat}}.
\end{figure}

Unfortunately, the He/H$_2$ ratio for Saturn is far from certain, as summarized
in Fig.\ \ref{Fig:herat}. From Pioneer Saturn
infrared radiometer measurements combined with Pioneer radio occultation
data, \cite{Orton1980} derived a mole fraction of H$_2$ equivalent to
He/H$_2$ = 0.11$\pm$0.03.  Using essentially the same method applied to
Voyager observations \citep{Hanel1981Sci} inferred a hydrogen abundance
equivalent to He/H$_2$ = 0.0638$\pm$0.007, while a later Voyager
analysis by \cite{Conrath1984He} yielded a value of the H$_2$ mole fraction
equivalent to
He/H$_2$ = 0.038$\pm$0.026. These values derived by radio occultation
and thermal spectral comparisons became suspect following the Galileo
Probe in situ measurements of 0.157$\pm$0.003 for the ratio on Jupiter
\citep{VonZahn1998,Niemann1998}, which is 1.6$\times\sigma$ larger than the Voyager
inferred value for Jupiter of 0.11$\pm$0.03 \citep{Gautier1981}. To avoid
whatever systematic errors that might be affecting the occultation-thermal
spectral method, \cite{Conrath2000} derived the ratio from IRIS spectra
alone, resulting in a value of 0.135$\pm$0.024.  More recent results
using Cassini radio occultations in combination with Cassini CIRS spectra,
by \cite{Gautier2006}, have yielded a value near 0.08 according to \cite{Fouchet2009},
although no specific value (or uncertainty) appears in the Gautier et al. (2006) abstract.
We will later show that if uniformity of the late wake region is to be explained
by the absence of deep aerosols, the He/H$_2$ ratio needs to be in the
range depicted by the gray bar in Fig.\ \ref{Fig:herat}, which is within
the lower range of values previously determined.

\subsection{Estimating uncertainties in derived model parameters}

Uncertainty estimates in derived model parameters depend critically on
estimates of the instrumental and gas absorption model errors, neither
of which is very well understood.  Random errors in the VIMS
measurements are only a minor contributor except at very low signal
levels.  Much more important are systematic effects arising from
instrument calibration errors, wavelength scale errors, and
uncertainty in gas and particulate absorption models. For example, we
have seen VIMS calibration changes over the years by as much as 20\%
in the 3-\mum region of the spectrum and by similar amounts due to
differences in flat field corrections. Presumably current calibration
errors have been reduced over time, but uncertainty estimates for the
current calibrations are not well defined.  Wavelength scale changes
have already been noted in a previous section, and there is some
uncertainty associated with attempts to correct for the known
wavelength shifts.

Following \cite{Sro2013gws}, we tried to account for these various
effects in a semi-realistic way using the following procedure.
Initial model fits were used to establish a rough characterization of
the vertical opacity structure of the atmosphere.  Model spectra were
then calculated for the rough model, and for two perturbations: (1)
optical depths offset by 0.01 and (2) optical depths increased by
10\%.  The resulting I/F differences were then root sum squared with
I/F errors due to wavelength uncertainties, and with an I/F offset
uncertainty of 5$\times 10^{-4}$ and a measurement/relative
calibration uncertainty assumed to be 1\%.  The I/F error associated
with wavelength uncertainty was calculated from the derivative of the
I/F spectrum with respect to wavelength times an estimated wavelength
uncertainty of 0.002 \mumx, which is 1/8 of a VIMS line width. An
alternative and simpler model computed I/F error as the root sum of squares of
an offset error of 0.005 in I/F and a fractional signal error of
6\%. The second model led to roughly the same fit characteristics as
the first on test cases. Thus we used the simpler model in most
calculations.  For comparison we note that \cite{Fletcher2011vims}
used a much more conservative error model for VIMS thermal spectra,
consisting of the greater of 12\% of radiance or 12\% of mean radiance
over the 4.6 -- 5.1 \mumx.  Our model results in smaller error
predictions, yet in most cases we are able to achieve fits with \chisq values
between 1 and 2 times the number of degrees of freedom (N$_F$ =  number
of fitted points minus the number of fitted parameters). To
approximately correct for bad error estimates and incomplete physics,
we re-scaled our \chisq values by the factor $N_F/\chi^2_{MIN}$ before
finding confidence limits.

\section{Spectrally constrained cloud structure}

We are able to accurately model the VIMS spectra with a relatively
simple cloud structure consisting of two main cloud layers.  The top
cloud is parameterized as a conservative cloud, using spherical Mie
particles with refractive index n = 1.4+0i, a top pressure ($p2t$), a bottom pressure
($p2$), a particle radius ($r2$), and an optical depth ($od2$). We
assume a gamma size distribution \citep{Hansen1974} with a fixed
variance of 0.1.  Although these particles are likely not spherical,
this choice of spherical particles is a convenient way to parameterize
wavelength dependence and is a reasonable approximation for small
enough size parameters \citep{Fry2014DPS}.  The bottom cloud is a
modeled as a sheet cloud of Henyey-Greenstein particles with
arbitrarily chosen single scattering albedo of $\varpi$ = 0.95 and
asymmetry parameter of g = 0 (these cannot be independently
constrained due to the significant optical depth of the overlying
cloud).  The lower cloud adjustable parameters are its optical depth
($odm$) and pressure ($pm$).  We did not attempt to constraint the
wavelength dependence of this cloud as it is mainly constrained by its
effects in the 5-\mum window.  Its inferred optical depth is a strong
function of its assumed single-scattering albedo.  This arises because
its effectiveness at blocking thermal emission is low with a high
single-scattering albedo, requiring more optical depth than
if the cloud is more absorbing. Because the observations are not
sensitive to the thickness of this cloud, we chose the simplification
of removing that parameter with the sheet cloud assumption (we made
the cloud top pressure 0.995 $\times$ the bottom pressure).  There is
also an optically thin haze layer of sub-micron particles (effective
radius = $r1$ with a similar gamma size distribution) at pressure $p1$
with optical depth $od1$. This layer is needed to fit the low but
non-zero I/F values in spectral regions with strong gas absorption.
For fits including reflected sunlight at modest incidence angles, we
find $r1$ = 0.14 \mum and $od1$ = 0.01 -- 0.02 work well, but these
parameters are not of much physical significance because the effect of
this layer is at the level of the VIMS offset uncertainty for the
observing conditions of our chosen spectral data.

\subsection{Sensitivity of model spectra to model parameters}

\begin{figure*}\centering
\includegraphics[width=6in]{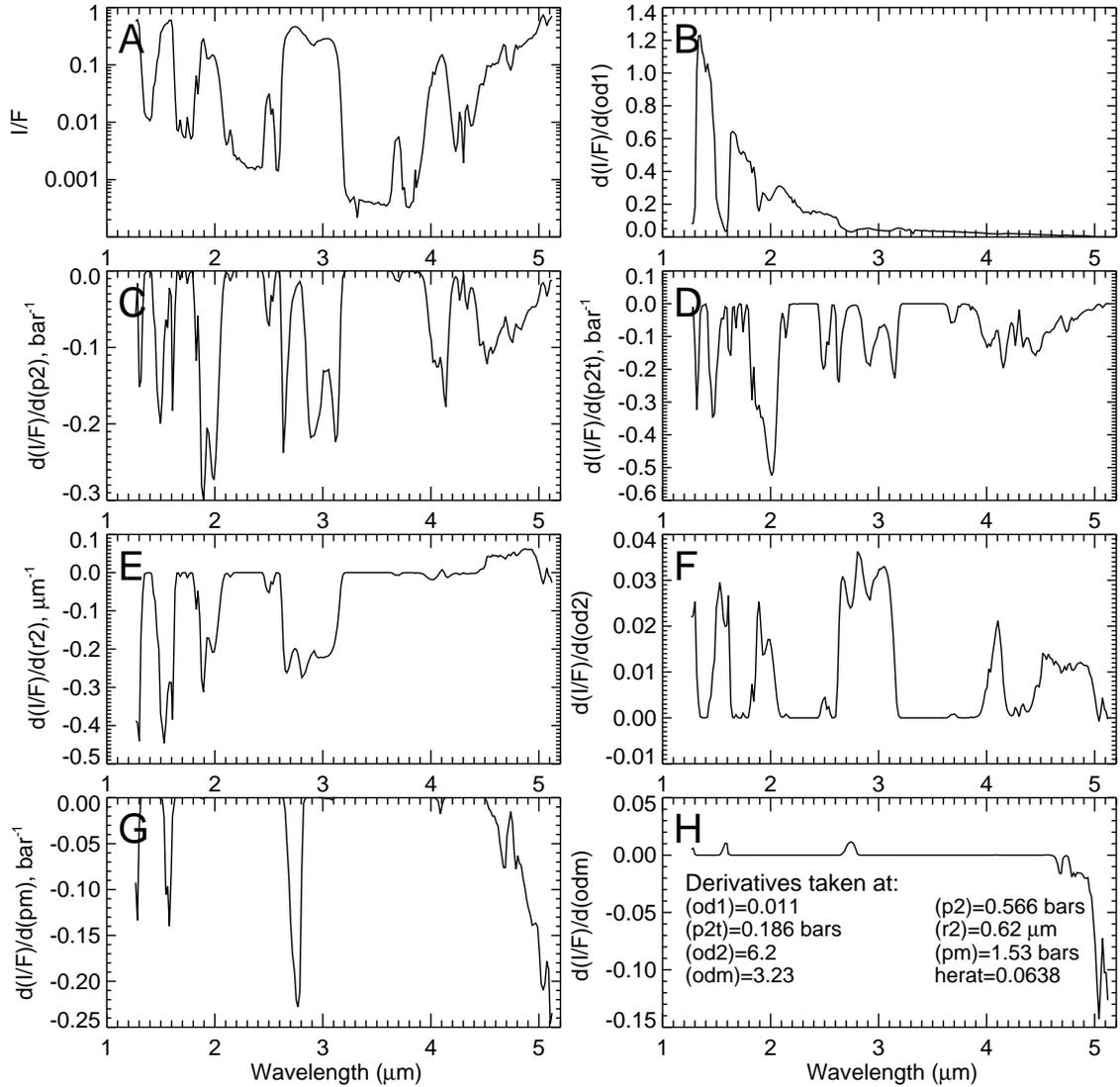}
\caption{Model I/F spectrum (upper left) and spectral derivatives of
  I/F with respect to seven key model parameters described in the
  text.  Panel H provides values of the parameters at which the
  derivatives were taken.
\label{Fig:deriv}}
\end{figure*}

The seven adjustable parameters we used are listed in Table
\ref{Tbl:paramlist}, along with the parameters that are either fixed
or manually altered, but not part of the Levenberg-Marquardt
non-linear fitting process.  For a pre-storm spectrum at 35\deg N
planetocentric latitude, we show in Fig.\ \ref{Fig:deriv} the
derivative of the model spectrum with respect to each of the normally
adjusted parameters. These derivative spectra are distinctly
different, although there are some significant correlations between
some of the parameters as listed in Table\ \ref{Tbl:corr}.  For
example, because both $p2t$ and $p2$ produce spectrally similar
decreases in the I/F spectrum, during the fitting process, these tend
to move in opposite directions to maintain a more constant effective
pressure for the layer.  This correlation could have been suppressed
by using an alternate parameterization in which mean pressure and
pressure thickness were the adjustable parameters. There are also
strong correlations between optical depth and cloud boundaries for
layer 2 for reflection spectra.  This arises because moving the top
pressure upward makes the cloud brighter, which needs to be
compensated for by reducing its optical thickness. Similarly, moving
the bottom downward makes the cloud darker, requiring a compensating
increase in optical depth.  None of the correlations have prevented
reasonably good constraints on particle parameters, as can be seen
from the fit results to follow.

Another important characteristic to take note of in
Fig.\ \ref{Fig:deriv} is that some parameters have vastly more important
effects on the part of the spectrum dominated by thermal emission
($\lambda > 4.5$ \mumx) relative to their effects on the part
dominated by reflected sunlight.  For example, the derivative with
respect to the optical depth of the deeper cloud layer (panel H) has
essentially no effect at most solar-dominated wavelengths, with the
main exception being near 2.7 \mumx, where the atmosphere is
sufficiently transparent that the light reflected by that deeper cloud
makes a small positive contribution.  Its effect at thermal
wavelengths is negative as a result of the absorption it
provides. Another dramatic spectral difference in effects can be seen in
panel B, where the derivative with respect to the optical depth of the
main upper cloud has only a tiny effect at wavelengths of thermal
emission, a result of the conservative nature of the cloud
particles.

\begin{table}\centering
\caption{Summary of cloud model parameters used in spectral calculations.}
\vspace{0.15in}
\begin{tabular}{r l l }
\hline
$p1$, bars & stratospheric haze pressure & 0.002\\
$r1$, \mum & stratospheric haze particle radius & 0.14 \mum \\
$od1$ & stratospheric haze optical depth & adjustable\\
$p2t$, bars & top of upper cloud & adjustable\\
$p2$, bars & bottom of upper cloud & adjustable\\
$r2$, \mum & radius of upper cloud particles & adjustable\\
$od2$ & optical depth of upper cloud at 2 \mum & adjustable\\
$pm$, bars & pressure of lower sheet cloud & adjustable\\
$odm$ & optical depth of lower sheet cloud & adjustable\\
$\varpi$ & single-scattering albedo of lower cloud & 0.95\\
g & asymmetry parameter of lower cloud & 0 (=isotropic) \\
H$_c$/H$_g$ & cloud particle to gas scale height ratio & 1.0 \\
\hline
\end{tabular}
\label{Tbl:paramlist}
\end{table}

\begin{table}\centering
\caption{Correlations between fitted parameters listed in Table\ \ref{Tbl:paramlist}.}
\vspace{0.15in}
\begin{tabular}{c c c c c c c c }
        & $p2$ & $p2t$ & $r2$   & $pm$ & $od1$   & $od2$  & $odm$   \\
\hline
  $p2$ & 1.000  &-0.915  & 0.096& 0.505  & -0.057 & 0.873 & -0.317  \\
 $p2t$ &-0.915  & 1.000 & -0.207& -0.372 & 0.134  & -0.706 & 0.227  \\
   $r2$ & 0.096  & -0.207 & 1.000 & 0.129  & 0.068  & 0.259 &  0.011 \\
 $pm$ & 0.505   &-0.372 & 0.129 & 1.000  & -0.003 & 0.659 & -0.719  \\
   $od1$ &-0.057  & 0.134  & 0.068& -0.003 & 1.000  &-0.015 &  0.008  \\ 
   $od2$ & 0.873  & -0.706  & 0.259& 0.659  & -0.015 & 1.000 & -0.389 \\
    $odm$ &-0.317  & 0.227  & 0.011&-0.719  & 0.008  &-0.389 &  1.000 \\
\hline
\end{tabular}
\label{Tbl:corr}
\end{table}

\subsection{Fit results for He/H$_2$ = 0.064}

\begin{figure*}\centering
\includegraphics[width=5.25in]{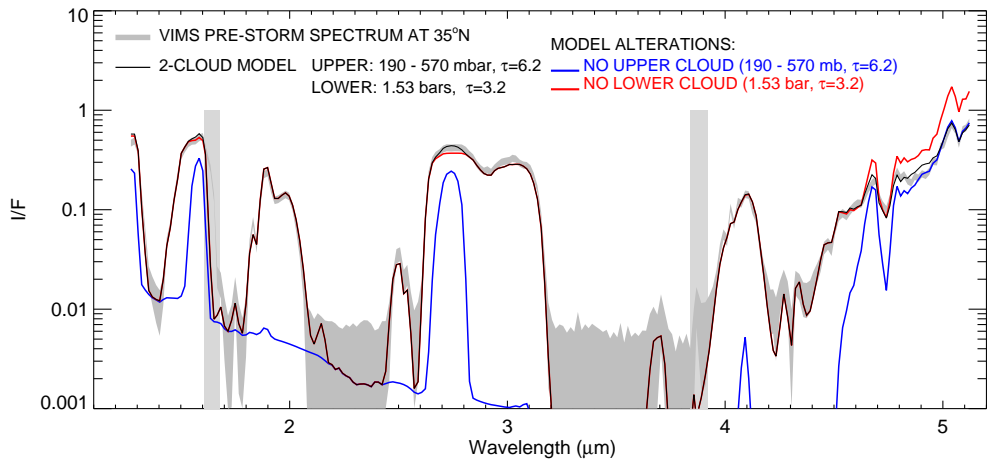}
\caption{February 2011 VIMS spectrum upstream of the Great Storm (gray),
model fit (black solid line), and spectra obtained by removing the upper
cloud only (blue) and the lower cloud only (red). Because it consists
of non-absorbing particles, the upper cloud does very little to attenuate
 the thermal emission.  The measured spectrum was extracted
from location A in Fig.\ \ref{Fig:color7}. The vertical light gray bars
  indicate regions where the VIMS calibration is unusable due to effects
  of order-sorting filter joints. \label{Fig:upfit}}  
\end{figure*}

\begin{figure*}\centering
\hspace{3.in}\includegraphics[width=3.in]{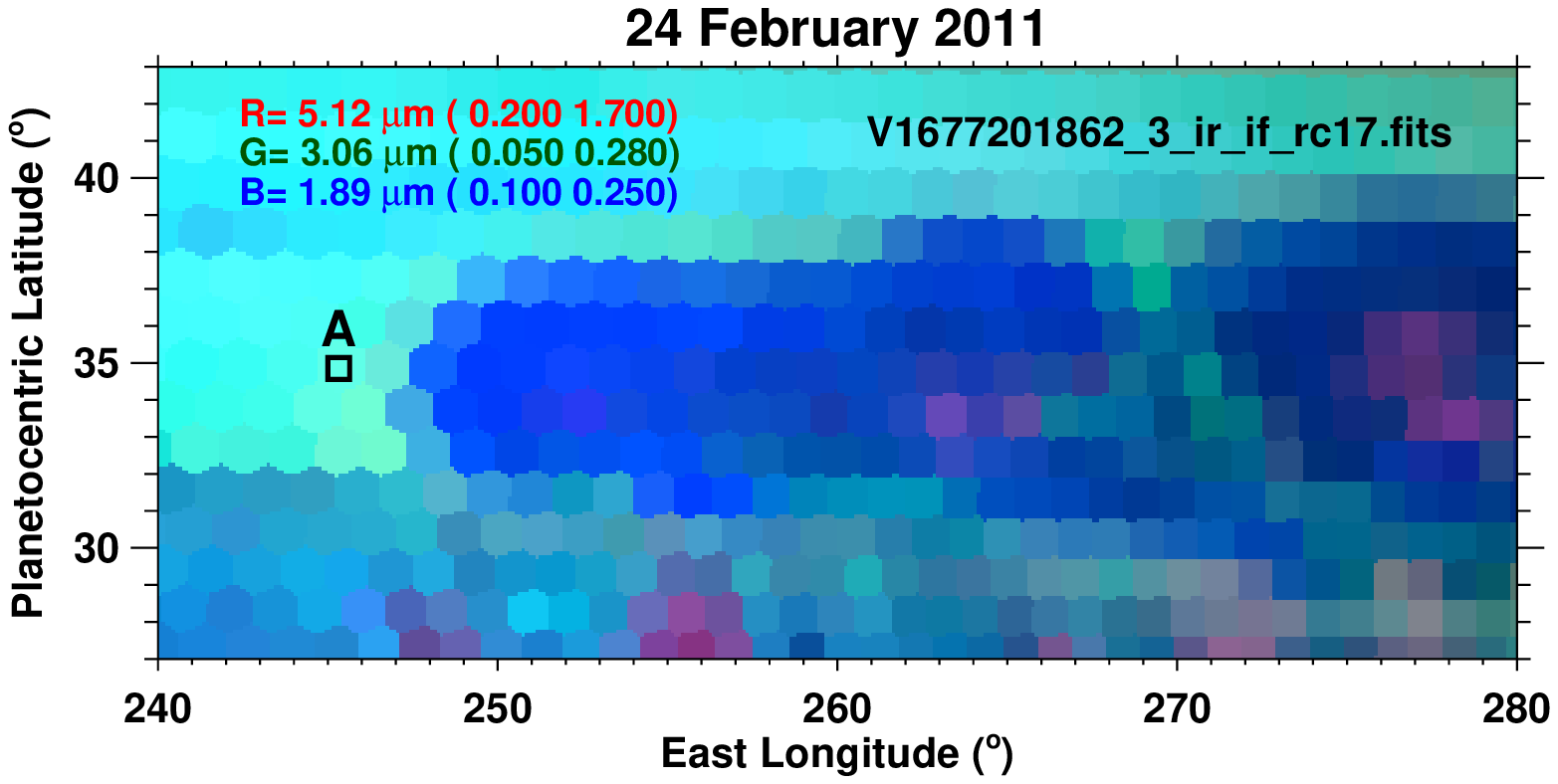}
\includegraphics[width=3.in]{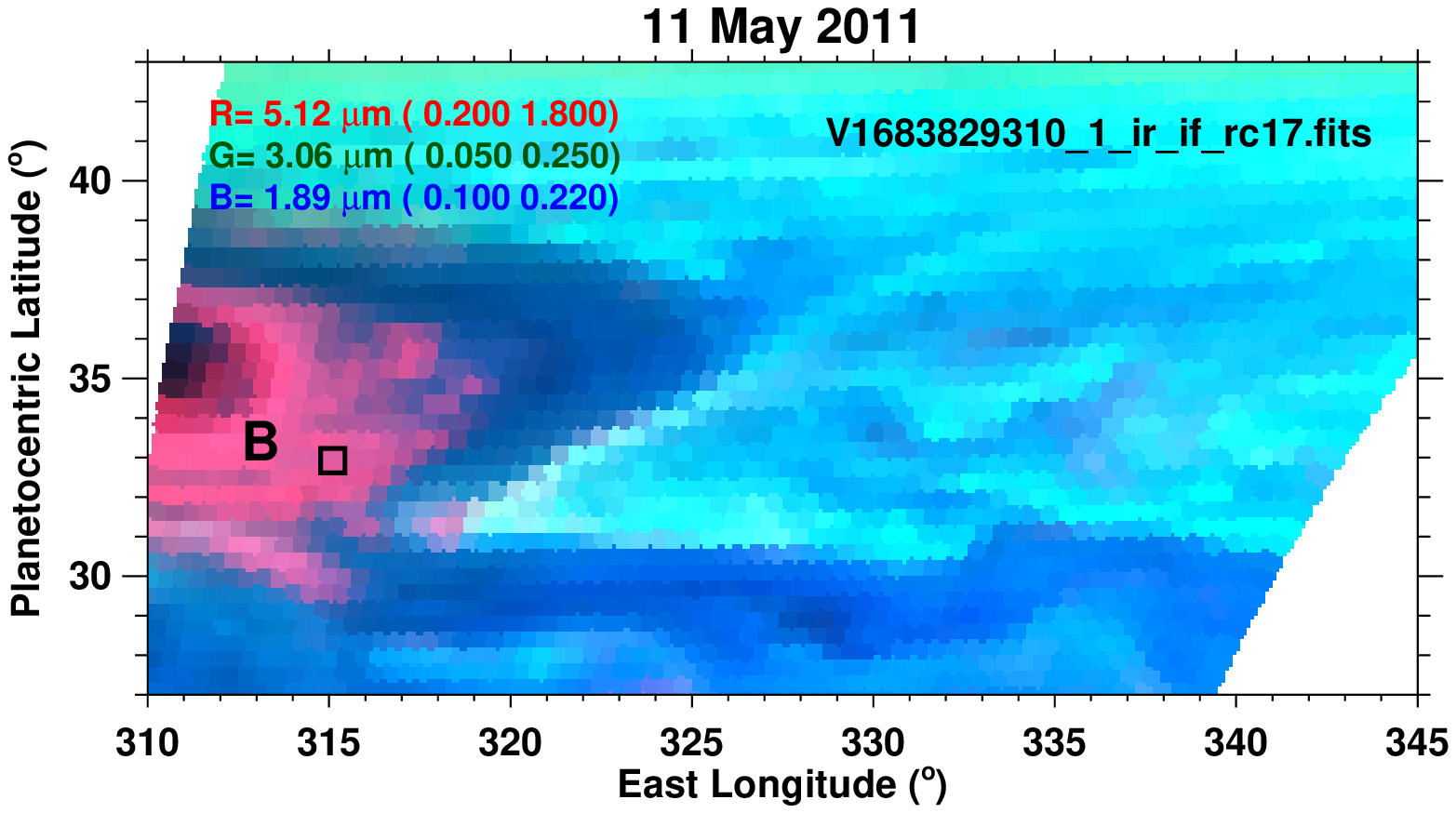}
\includegraphics[width=3.in]{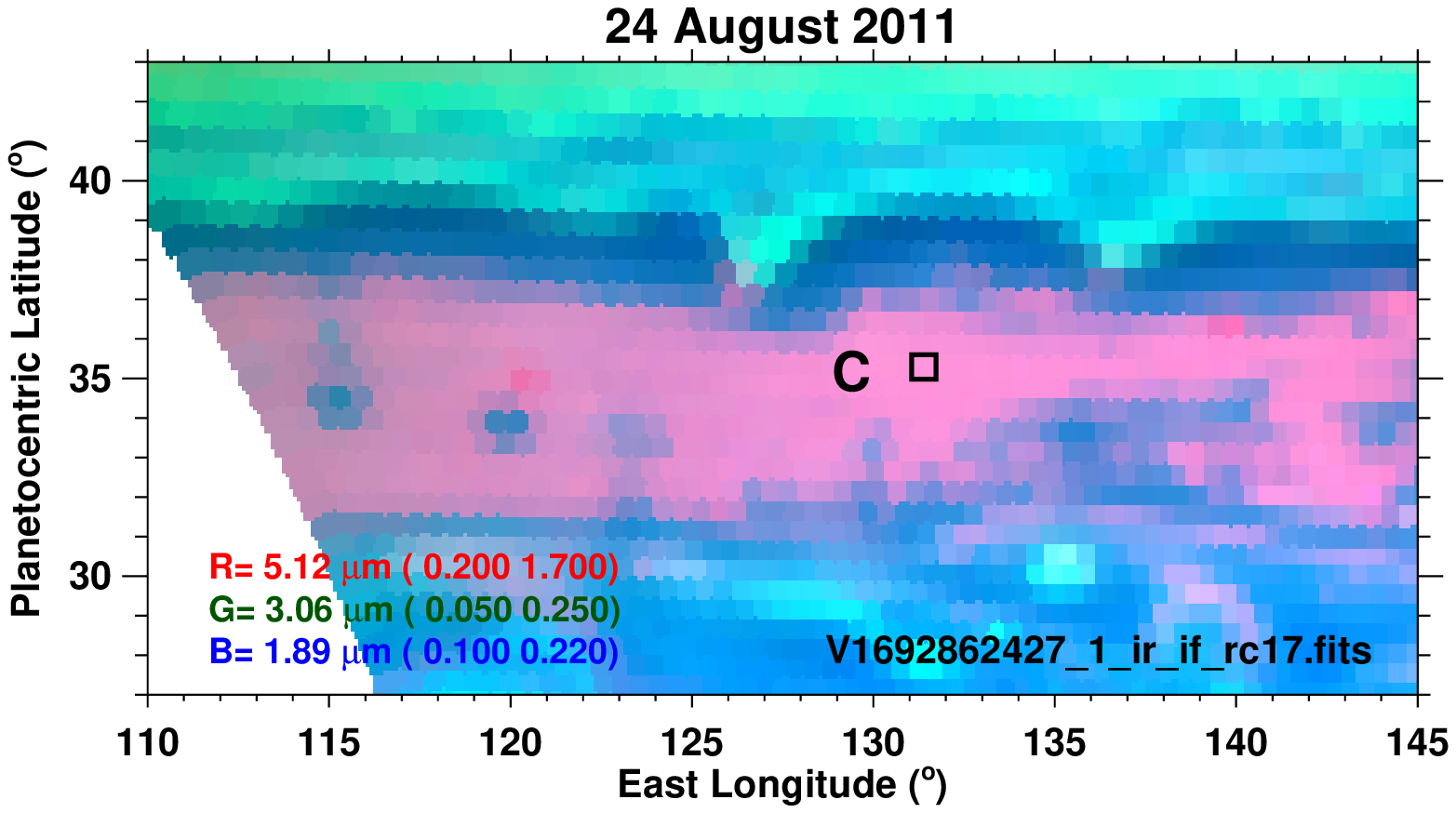}
\includegraphics[width=3.in]{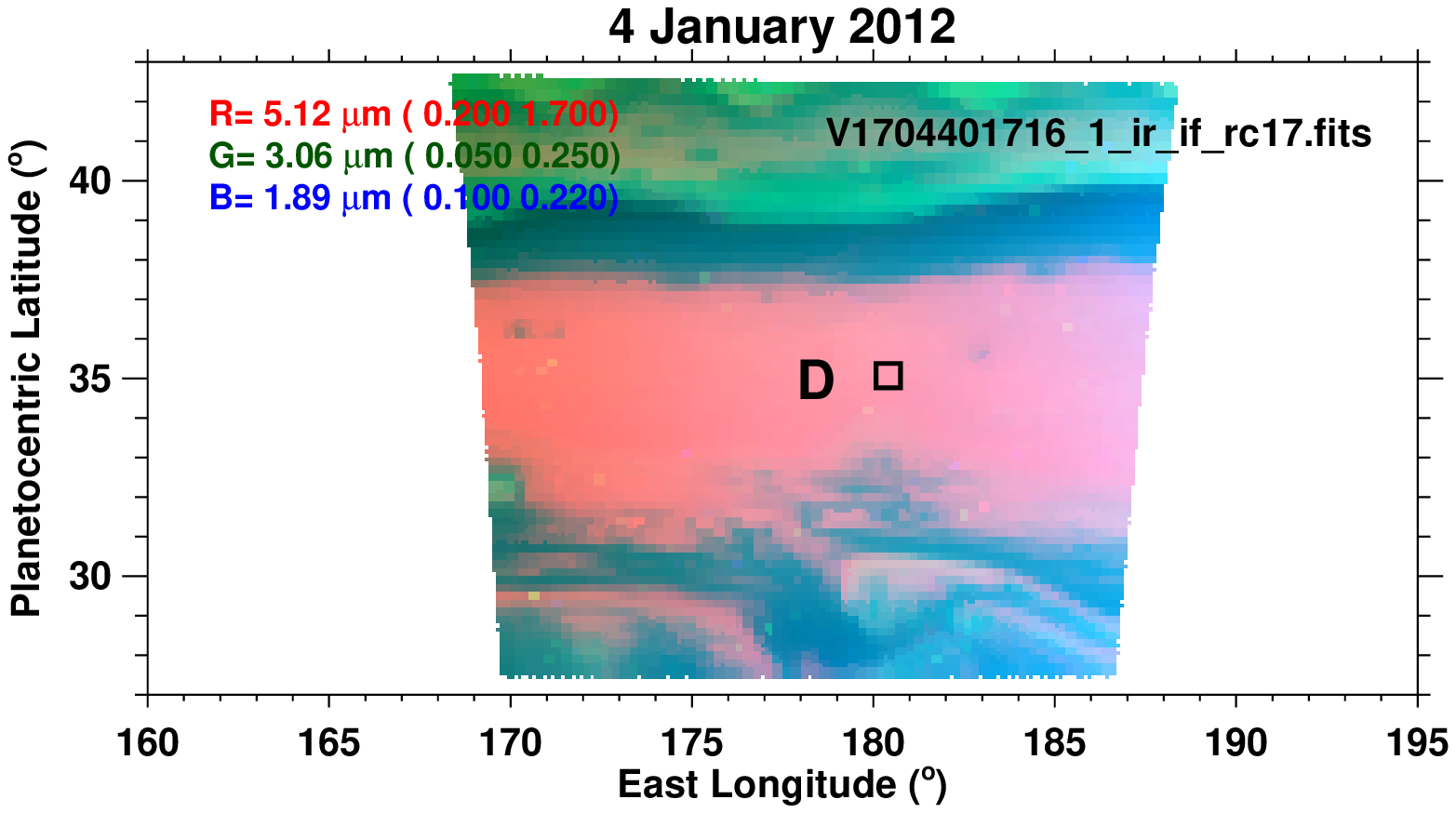}
\includegraphics[width=3.in]{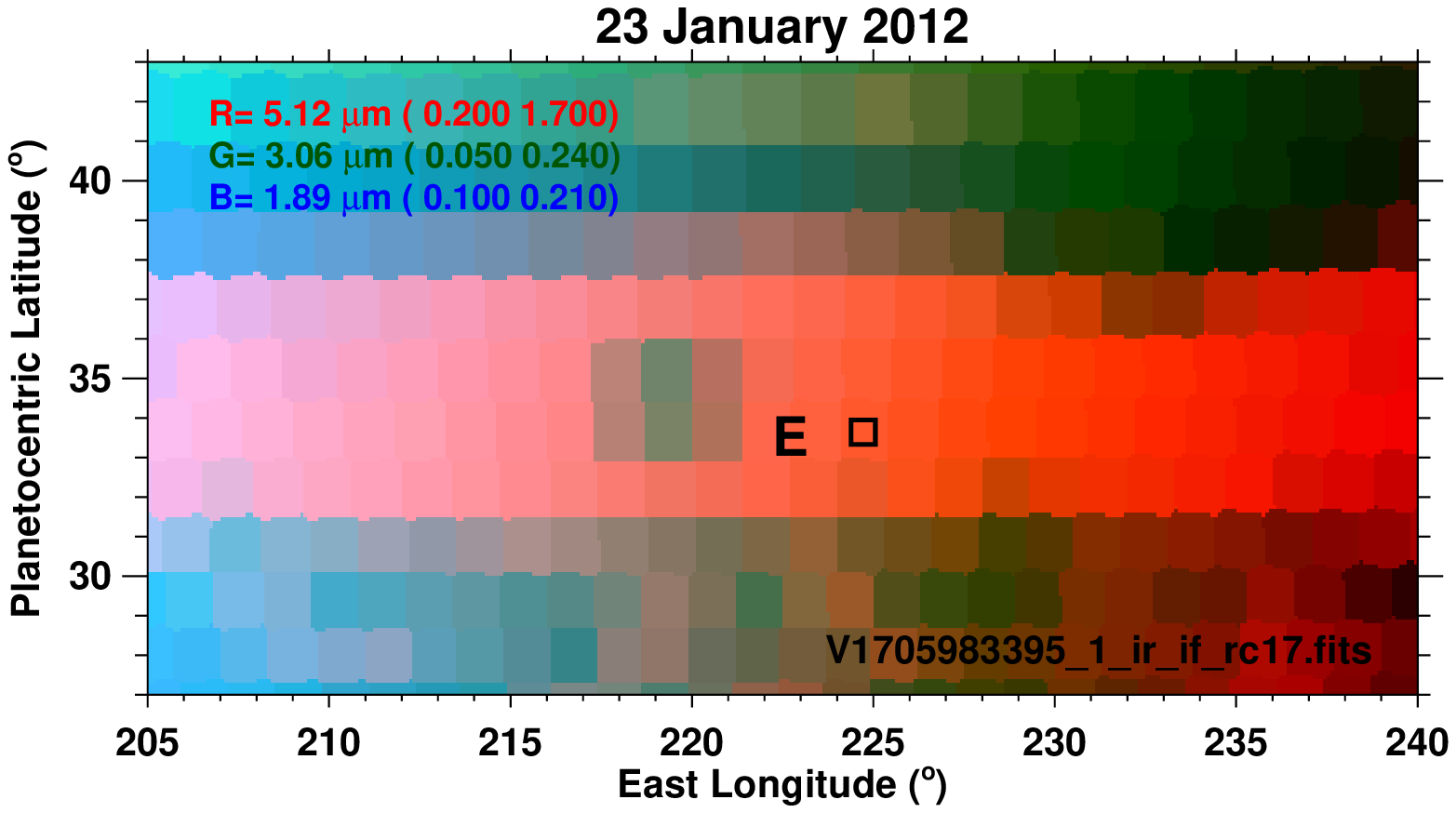}
\includegraphics[width=3.in]{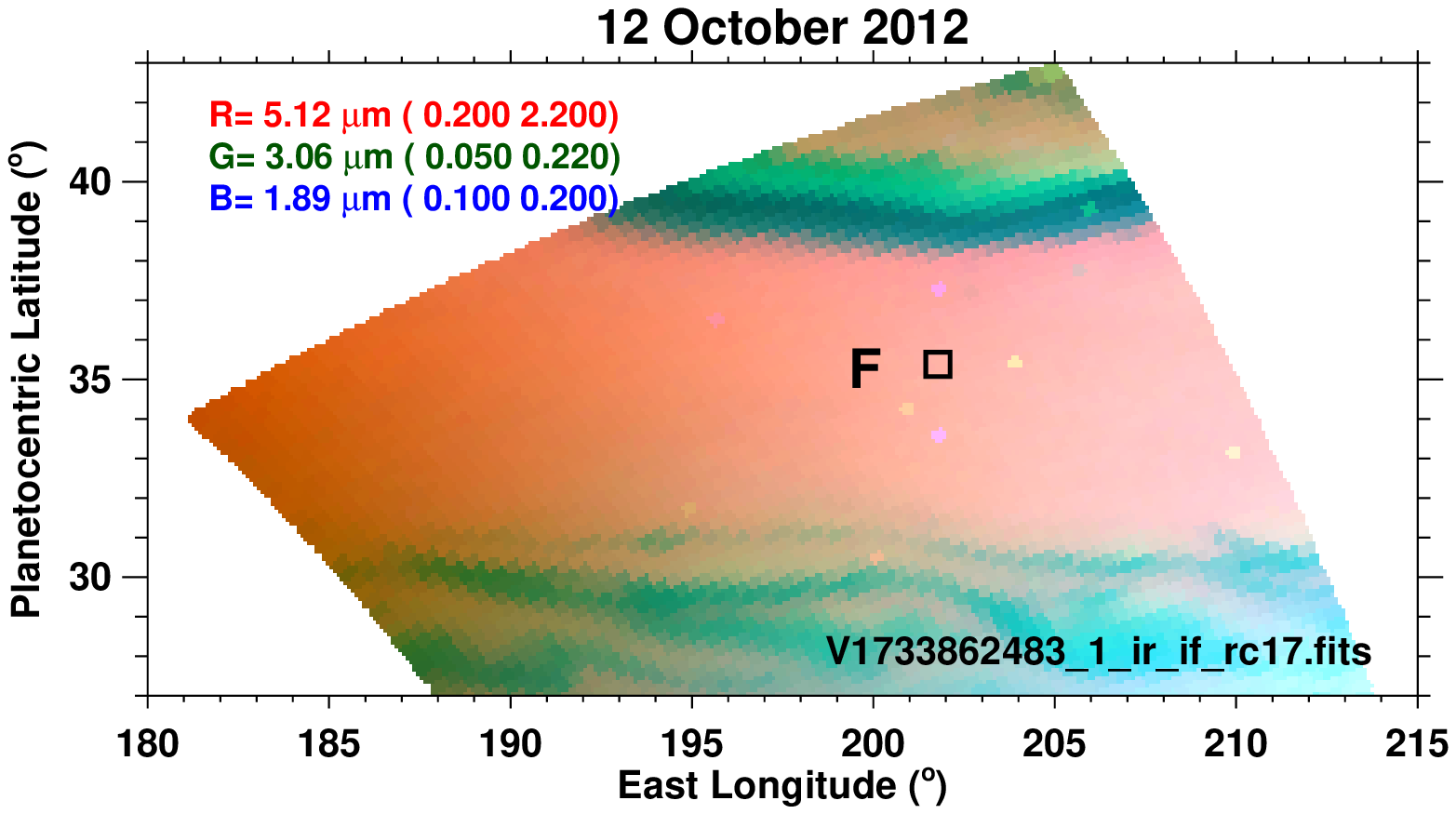}
\includegraphics[width=3.in]{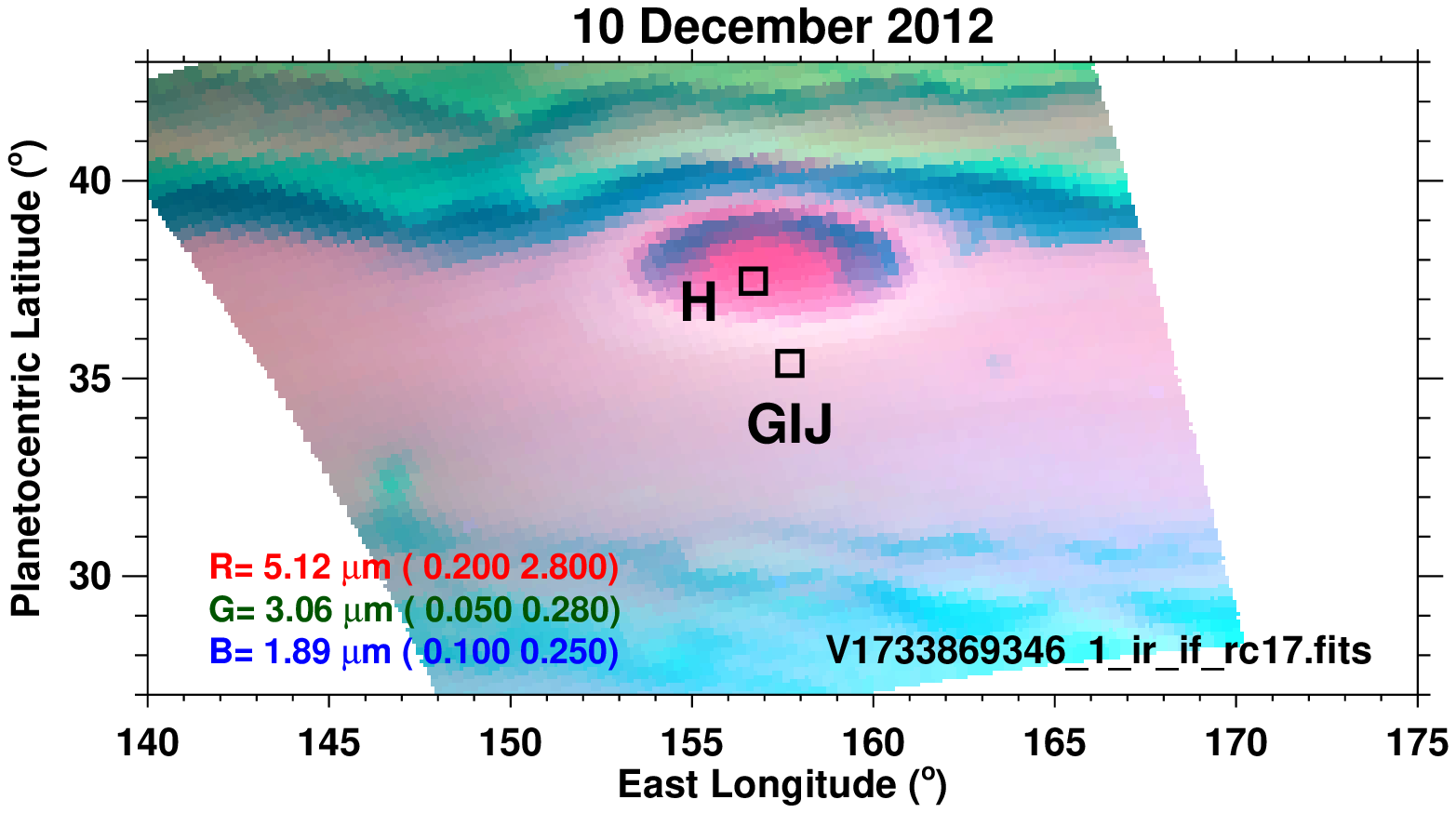}
\caption{Color composite VIMS image of the Great Storm on 24 February
  2011 (top) and selected images of the wake region on six dates from
  11 May 2011 through 10 December 2012.  R, G, and B channels are
  assigned to wavelengths of 5.12 \mumx, 3.06 \mumx, and 1.89 \mumx,
  with stretches given in the legends. In each image the location(s)
  of spectral samples are marked by black squares and labeled A-J, for
  which fit results are presented in Table\ \ref{Tbl:fittab}. In May
  2011 and December 2012 images, the oval feature is the anticyclonic
  vortex labeled as AV in Fig.\ \ref{Fig:5mos}, which also provides a
  larger context for all but the October sample.}
\label{Fig:color7}
\end{figure*}

A sample fit to an upstream spectrum is provided in
Fig.\ \ref{Fig:upfit}, with fit parameters, uncertainties, and fit
quality provided in the first row of Table\ \ref{Tbl:fittab}.  This
spectrum, from location A in Fig.\ \ref{Fig:color7},
has a relatively low I/F in the 4.6 -- 5.12 \mum region and thus the
model requires a lower cloud of significant absorption optical depth
to limit thermal emission.  As can be seen in Fig.\ \ref{Fig:upfit}
from what happens to the model spectrum when the upper cloud is
removed, the overlying cloud of conservative particles provides little
attenuation. The assumed lower cloud properties ($\varpi$ = 0.95, g =
0) do provide sufficient attenuation with a modest optical depth of
3.2.  The pressure of this cloud is constrained by its role in shaping
the thermal emission spectrum and in adding reflectivity in the
continuum regions at shorter wavelengths, especially near 2.8 \mum
(see Fig.\ \ref{Fig:deriv}G).  

\begin{figure*}\centering
\includegraphics[width=5.25in]{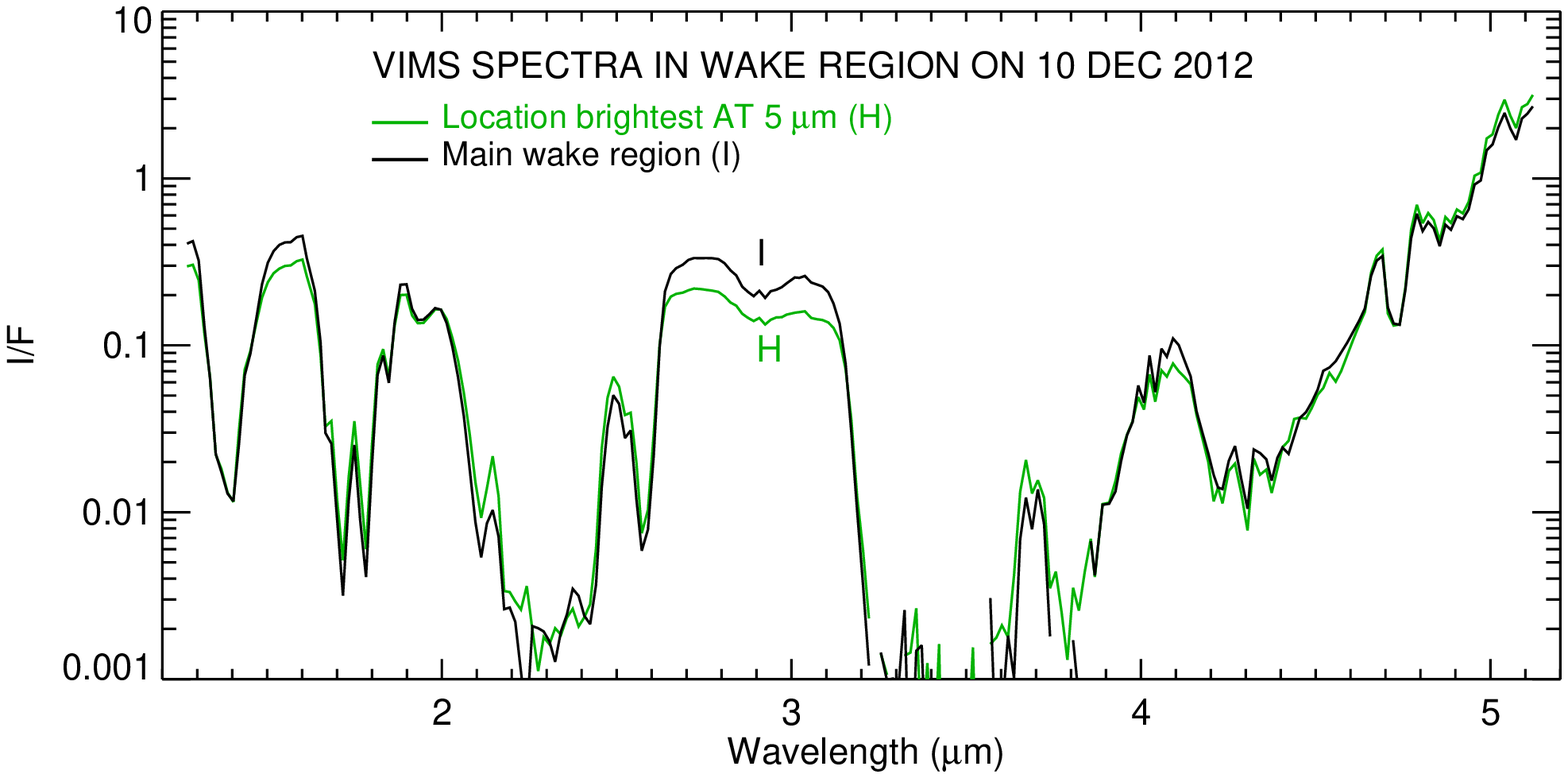}
\caption{Comparison of 10 December 2012 spectra obtained in the
  central wake region (location I in Fig.\ \ref{Fig:color7}) and the
  5-\mum brightest region in the center of the anticyclonic vortex
  (location H).  Note that the latter is darker at continuum
  wavelengths below 4.5 \mum where reflected sunlight dominates,
  indicating reduced upper cloud optical depth.\label{Fig:spec12}}
\end{figure*}

\begin{figure*}\centering
\includegraphics[width=5.25in]{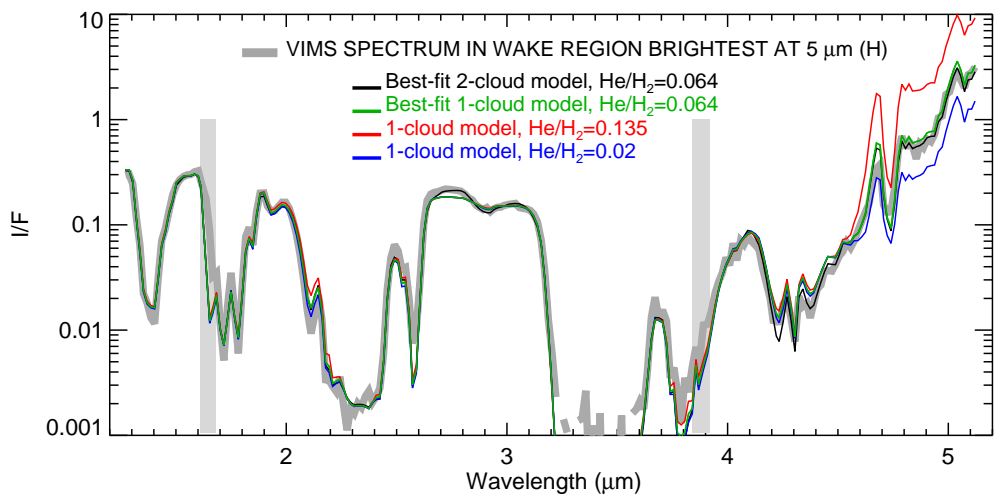}
\caption{A spectrum measured in the core of the anticyclone (gray
  thick line) and alternative fits using a two-cloud model (thin black
  line) and a one-cloud model (green line). Also shown are spectra
  computed from the one-cloud model parameters, but with temperature
  profiles for \herat = 0.02 (blue) and \herat = 0.135 (red).  These
  are discussed in Sections 6.1 and 6.2.  The vertical light gray bars
  indicate regions where the VIMS calibration is  unusable due to effects
  of order-sorting filter joints. \label{Fig:brightfit}}
\end{figure*}

\begin{figure*}\centering
\includegraphics[width=5.25in]{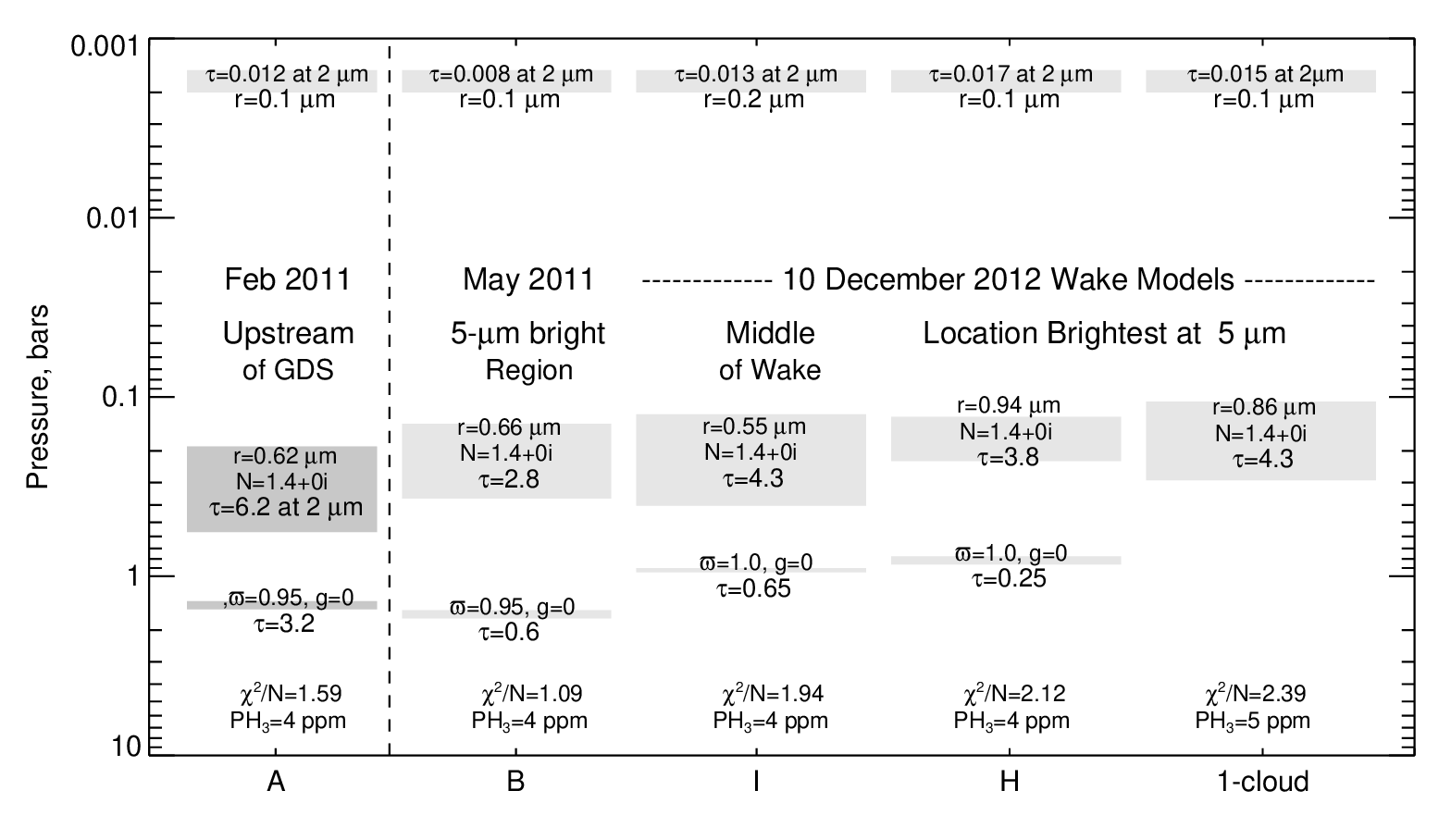}
\caption{Comparison of the two-cloud model (A) for the region upstream
  of the Great Storm with two-cloud models (B, I, and H) for selected
  5-\mum bright wake regions, and with the best-fit one-cloud model from
  Fig.\ \ref{Fig:corefits}, which is discussed in Sec. 5.4.  The above
  letter labels correspond to those given in Table\ \ref{Tbl:fittab}
  and Fig.\ \ref{Fig:color7}. \label{Fig:cartoon}}
\end{figure*}

Spectra from the 10 December 2012 observations (locations H and I in
Fig.\ \ref{Fig:color7}) are compared in Fig.\ \ref{Fig:spec12}, with
the spectrum from the main wake region (H) shown as green and that
from the AV core (I) shown as black.  There is little difference
between them in the thermal emission part of the spectrum ($\lambda >
4.5$ \mumx). The more significant difference is at continuum
wavelengths that are controlled by scattered light from the upper
cloud.  Clearly the upper cloud is optically thinner over the
anticyclone, although it still makes a significant contribution to the
spectrum.  Sample fits to the H spectrum, the region brightest at 5
\mum on 10 December 2012, are shown as black and green curves in
Fig.\ \ref{Fig:brightfit} (other model spectra shown there are
discussed in Sections 6.1 and 6.2). The corresponding cloud structures
are displayed in Fig.\ \ref{Fig:cartoon}.  For \herat = 0.064, and
\pht VMR = 4 ppm (with a uniform mixing ratio for p $\ge$ 0.55 bars
and falling off with a gas/pressure scale height ratio of 0.5 for p
$<$ 0.55 bars), the two-cloud model provides a better fit than the
one-cloud model (\chisq/N$_\mathrm{F}$ = 2.12 vs. 2.39).  The
one-cloud model does not provide a high enough I/F near 2.75 \mumx,
which indicates the need for an additional cloud layer, but that cloud
does not need to be an absorbing cloud and its inferred pressure (from
the two-cloud model) is near 800 mb, rather than at the 1.5 -- 1.6 bar
level seen before and during the early stages of the
clearing. Clearly, this cloud does not need to play a significant role
in attenuating thermal emission.

\begin{figure}[!hbt]\centering
\includegraphics[width=3.5in]{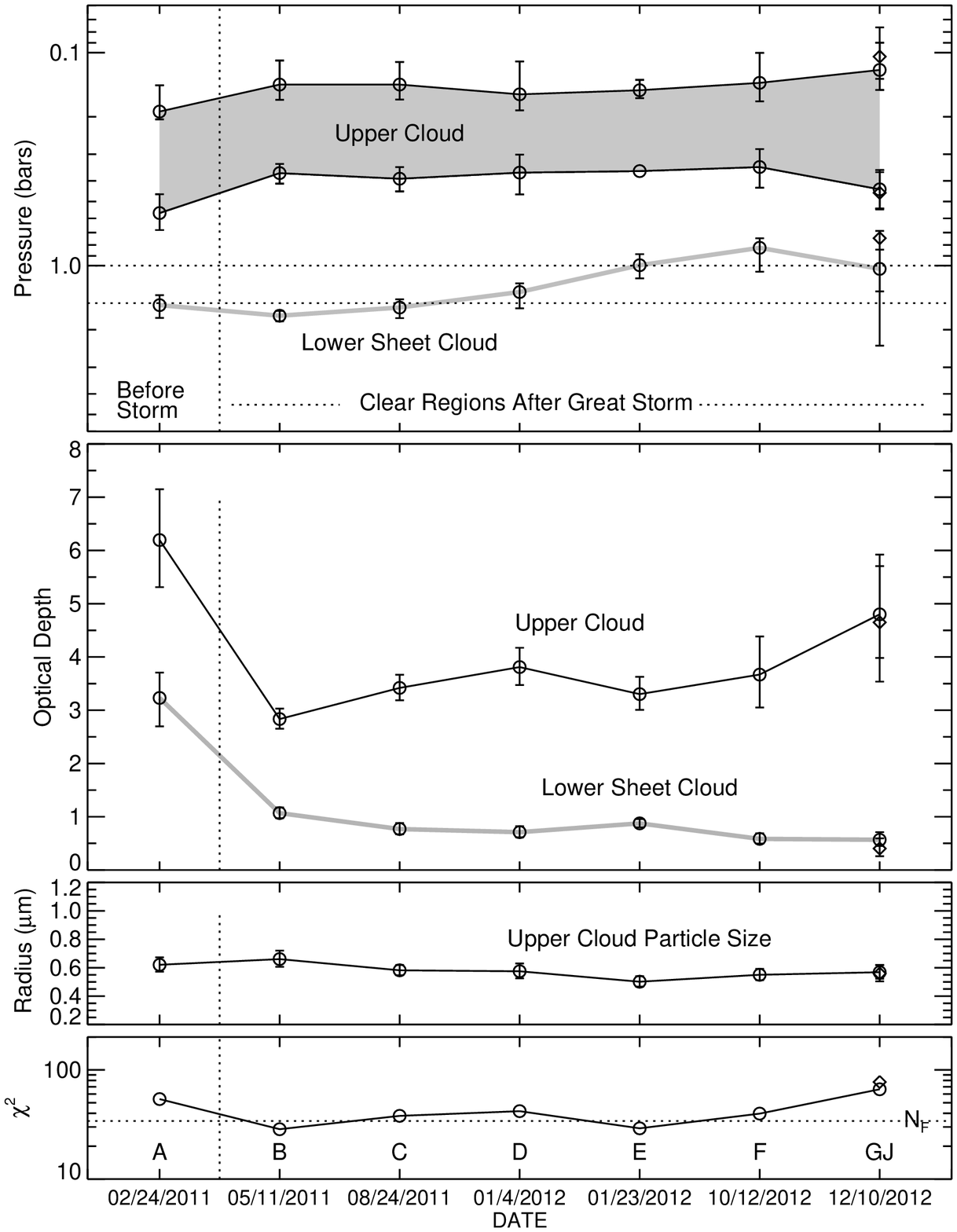}
\caption{Main fit parameters for vertical cloud structure models
  constrained by reflected solar and thermal spectra from ``clear''
  (5-\mum bright) regions, as a function of time. The initially low
  upper cloud opacity seems to be slowly returning to the optical
  depth it had prior to the Great Storm, although the larger
  uncertainty in the later points leaves open the
  possibility that the trend is not real.  The letter labels in the
  bottom panel refer to fits in Table\ \ref{Tbl:fittab} and locations
  in Fig.\ \ref{Fig:color7}.  Points plotted with diamond symbols are
  for fit J, for which the \pht profile was scaled to make the deep
  mixing ratio equal 5 ppm.  This slightly reduces the best-fit
  optical depth of the lower cloud and moves it to lower pressures,
  towards the bottom of the upper cloud.  The worsening fit quality at
  later times may be due to drifts in the VIMS spectral
  scale.\label{Fig:fitsum}}
\end{figure}

Table \ref{Tbl:fittab} summarizes the main fit results from spectra
containing both solar and thermal contributions, which we found
provided better constraints on upper cloud structure than were
possible with just night-side thermal emission spectra. The observing
geometry for each spectrum we fit is given in Table\ \ref{Tbl:geom}.  Note that these
results are for \herat = 0.064.  The
main parameters are plotted versus time in Fig.\ \ref{Fig:fitsum}.
Letter labels in this figure and the referenced tables refer to locations
given in Fig.\ \ref{Fig:color7}.
The earliest observation we fit was obtained on 24 February 2011, from
region A in that figure, immediately upstream of the Great
Storm, before it was affected by the spreading wake.  This represents
the cloud structure before the ``clearing-out'' process in the wake
took effect.  The earliest observation we fit inside the wake was
obtained on 11 May 2011 in the 5-\mum bright region in the vicinity of
the large anticyclone (location B in Fig.\ \ref{Fig:color7}), before
it was overtaken by the Great Storm itself, which happened in June
2011.  The anticyclone survived the encounter, and has remained
through at least 19 August 2015 \citep{Momary2015}. By January 2012,
the clearing process produced high emitted radiances that extended all
the way around the planet, and by December 2012 it achieved a high
degree of longitudinal uniformity, as well as a moderately high
latitudinal uniformity within planetocentric latitudinal boundaries
from 32\deg N to 39\deg N (see Fig.\ \ref{Fig:5mos}).

\begin{table*}\centering
\caption{Fit results for 5-\mum bright regions in the wake of Saturn's Great Storm of 2010-2011.}
\vspace{0.015in}
\begin{tabular}{c c c c c c c c c c c}
\hline
   &$p2t$ & $p2$ & $r2$      &       & $pm$  &       &        & PC & East &  \\[0.1ex]
ID &(mbar) & (mbar) & (\mum) & $od2$ & (bar) & $odm$ & $\chi^2$ & Lat.    & Lon. &{\footnotesize MM/DD/YYYY} \\[0.1ex]
\hline
\\[-2ex]
A &189$^{+ 16}_{- 46}$ & 566$^{+113}_{-103}$ &  0.62$^{+0.05}_{-0.05}$ & 6.20$^{+0.95}_{-0.88}$ & 1.53$^{+0.23}_{-0.16}$ & 3.23$^{+0.47}_{-0.53}$ &  54.0 &  34.9\deg &  245.3\deg &  02/24/2011\\[0.5ex]
\hline
\\[-2ex]
B &141$^{+ 25}_{- 32}$ & 367$^{+ 44}_{- 34}$ &  0.66$^{+0.06}_{-0.05}$ & 2.83$^{+0.20}_{-0.18}$ & 1.72$^{+0.11}_{-0.09}$ & 1.07$^{+0.10}_{-0.10}$ &   28.6 &  32.9\deg &  315.1\deg & 05/11/2011\\[0.5ex]
C &141$^{+ 24}_{- 30}$ & 390$^{+ 57}_{- 45}$ &  0.58$^{+0.04}_{-0.03}$ & 3.42$^{+0.25}_{-0.23}$ & 1.57$^{+0.19}_{-0.13}$ & 0.77$^{+0.11}_{-0.10}$ &   37.9 &  35.3\deg &  131.4\deg & 08/24/2011\\[0.5ex]
D &157$^{+ 29}_{- 46}$ & 366$^{+ 97}_{- 64}$ &  0.57$^{+0.05}_{-0.05}$ & 3.81$^{+0.36}_{-0.34}$ & 1.33$^{+0.26}_{-0.12}$ & 0.71$^{+0.11}_{-0.10}$ &   41.9 &  35.1\deg &  180.4\deg &  01/04/2012\\[0.5ex]
E &150$^{+ 13}_{- 15}$ & 360$^{+  0}_{-  0}$ &  0.50$^{+0.04}_{-0.03}$ & 3.30$^{+0.32}_{-0.30}$ & 1.00$^{+0.15}_{-0.11}$ & 0.88$^{+0.08}_{-0.07}$ &   29.1 &  33.6\deg &  224.7\deg & 01/23/2012\\[0.5ex]
F &138$^{+ 30}_{- 38}$ & 344$^{+ 86}_{- 60}$ &  0.55$^{+0.04}_{-0.04}$ & 3.67$^{+0.72}_{-0.62}$ & 0.82$^{+0.24}_{-0.08}$ & 0.58$^{+0.10}_{-0.09}$ &   39.8 &  35.4\deg &  201.8\deg & 10/12/2012\\[0.5ex]
G &120$^{+ 29}_{- 30}$ & 438$^{+100}_{- 82}$ &  0.57$^{+0.05}_{-0.05}$ & 4.80$^{+0.90}_{-0.82}$ & 1.04$^{+0.29}_{-0.19}$ & 0.57$^{+0.14}_{-0.12}$ &   66.4 &  35.2\deg &  157.7\deg & 12/10/2012\\[0.5ex]
H &128$^{+ 41}_{- 51}$ & 227$^{+ 96}_{- 26}$ &  0.94$^{+0.13}_{-0.12}$ & 3.79$^{+0.53}_{-0.49}$ & 0.86$^{+0.44}_{-0.14}$ & 0.25$^{+0.12}_{-0.09}$ &   63.6 &  37.5\deg &  156.7\deg & 12/10/2012\\[0.5ex]
I&125$^{+ 29}_{- 32}$ & 403$^{+ 84}_{- 76}$ &  0.55$^{+0.08}_{-0.07}$ & 4.29$^{+0.82}_{-0.74}$ & 0.95$^{+0.20}_{-0.12}$ & 0.65$^{+0.14}_{-0.12}$ &   63.7 &  35.2\deg &  157.7\deg & 12/10/2012\\[0.5ex]
\hline
\hline
\\[-2ex]
J&104$^{+ 28}_{- 28}$ & 455$^{+ 89}_{- 90}$ &  0.56$^{+0.06}_{-0.06}$ & 4.65$^{+1.27}_{-1.11}$ & 0.74$^{+1.63}_{-0.06}$ & 0.40$^{+0.19}_{-0.14}$ &   77.4 &  35.2\deg &  157.7\deg & 12/10/2012\\[0.5ex]
\hline
\\[-2ex]
\end{tabular}\label{Tbl:fittab}

\noindent
\parbox{5.75in}{Note: the first fit (A) is for a region upstream of
  the Great Storm and undisturbed by the wake. The remaining fits are
  for regions inside the wake that exhibit high 5-\mum emission. The
  deep phosphine VMR was set to 4 ppm for fits above the double line
  (A-I) and the profile was scaled to produce a deep VMR of 5 ppm for
  fit J.  Fits H and I used $\varpi = 1$ for the lower cloud, while
  the rest used $\varpi = $ 0.95. Fit H is for the brightest 5-\mum
  region. Observing geometry is given in Table\ \ref{Tbl:geom}, and the
  locations from which fitted spectra were extracted are identified in
  Fig.\ \ref{Fig:color7}.}
\end{table*}

\begin{table}\centering
\caption{Observing geometry for spectral fits in Table\ \ref{Tbl:fittab}.}
\vspace{0.015in}
\begin{tabular}{c c c c c}
\hline
 &  Observer & Solar & Azimuth & Phase  \\[0.1ex]
ID &  zenith angle & zenith angle & angle & angle\\[0.1ex]
\hline
\\[-2ex]
   A&  51.31\deg &  28.98\deg &  98.52\deg &  51.81\deg \\[0.5ex]
   B&  39.13\deg &  24.00\deg & 139.64\deg &  23.14\deg \\[0.5ex]
   C&  40.66\deg &  34.16\deg & 154.67\deg &  15.37\deg \\[0.5ex]
   D&  40.59\deg &  53.21\deg &  98.19\deg &  56.04\deg \\[0.5ex]
   E&  35.82\deg &  68.82\deg &  87.66\deg &  74.10\deg \\[0.5ex]
   F&  27.81\deg &  55.41\deg & 146.17\deg &  36.41\deg \\[0.5ex]
   G&  16.37\deg &  39.72\deg & 132.68\deg &  31.25\deg \\[0.5ex]
   H&  15.75\deg &  40.84\deg & 138.91\deg &  31.25\deg \\[0.5ex]
   I&  16.37\deg &  39.72\deg & 132.68\deg &  31.25\deg \\[0.5ex]
   J&  16.37\deg &  39.72\deg & 132.68\deg &  31.25\deg \\[0.5ex]
\hline
\\[-2ex]
\end{tabular}\label{Tbl:geom}
\end{table}

\subsection{Evolution of cloud structure in the ``cleared'' regions}

Relative to the undisturbed cloud structure ahead of the storm, the
earliest bright region we sampled (B) had 70\% less upper-cloud
optical depth and lower boundary pressures, as well as a factor of 4
-- 5 less optical depth for the deeper cloud, with no difference in
pressure.  But as the wake clearing spread, the
lower cloud in the cleared regions rose in altitude to near the 1-bar
level.  This might mean that the deeper lower cloud just disappeared
and a previously insignificant layer became more noticeable.
Most of the fits were made with a deep \pht VMR of 4 ppm; with 5 ppm,
the needed lower cloud opacities decrease somewhat, raising the
possibility of a complete clearing of lower cloud particles.  It is
also conceivable that small reductions in temperature produce lower
emissions, which in turn would require less blocking of those
emissions and might also be consistent with a complete clearing of
lower cloud particles. A 1 K change in effective temperature would
produce about a 6\% change in the emitted radiance at 5 \mumx.
The effect of the \herat
ratio on the lower cloud properties, and how well spectra can be fit
without a lower cloud, are discussed further in the following section.

\subsection{Alternative models of cloud structure in the clearest regions }

At middle latitudes, the 5-\mum brightest region on Saturn seems to be
at the edge of the ``cleared-out'' band inside the core of what
appears to be the remnant of the anticyclonic vortex (AV) that formed
along with the Great Storm, but persisted for a much longer time
interval \citep{Sayanagi2013,Momary2014,Momary2015}.  VIMS spectral
observations from 10 December 2012, plotted in Fig. \ref{Fig:spec12},
show that this core region (H) is slightly brighter than the middle of
the bright band (I), possibly because there is less high-altitude
cloud optical depth to attenuate the thermal emission. At short
wavelengths this core region is relatively darker than the middle of
the band, which tends to support this speculation, although this
effect should be small if the upper cloud is truly conservative.
Spectral fits to the core region spectrum are shown in
Fig.\ \ref{Fig:corefits} for a range of \herat ratios.  Even this region of very
high 5-\mum brightness and low visible I/F is seen to have a significant
optical depth of overlying cloud particles, increasing from $\sim$3.5
to $\sim$5.5 for \herat ratios from 0.02 to 0.09.

\subsubsection{Fits of two-cloud models as a function of the \herat ratio}

When we model the structure using two clouds, we find that the upper
cloud layer extends from 150 mbar to 250 mbar and has an optical depth
near 4, with little dependence on the assumed He/H$_2$ ratio. Fit
quality is also maintained over a wide range of \herat ratios, mainly
by adjusting the \pht mixing ratio and the lower cloud optical depth
and pressure.  The inferred properties of the lower cloud in our
model vary significantly with the assumed He/H$_2$ ratio.  At large
values of that ratio, the sheet cloud is found near 1.8 bars, with an
optical depth of 1.6 to 2.4 for deep \pht mixing ratios from 6\tenmsix
to 4\tenmsixx, respectively.  As the \herat ratio decreases the
optical depth of the lower cloud also decreases, reaching zero at a
\herat ratio of 0.05 for a \pht VMR of 4\tenmsixx, at a \herat ratio
of 0.064\tenmsix for a \pht VMR of 5\tenmsixx, and at larger ratios
for larger values of the \pht mixing ratio.  Even though we found a
non-zero optical depth for the lower cloud at \herat = 0.064, that
cloud is not providing any significant attenuation of the thermal
emission at that point. In fact, if we make this lower cloud
conservative (setting $\varpi$ = 1.0), as for fit H in
Table\ \ref{Tbl:fittab}, the results is a better fit and an even
smaller optical depth.  Thus, at this value of the \herat ratio, there
is really no evidence for a deep absorbing cloud, although one is
clearly required at larger \herat ratios (assuming fixed trace gas
profiles). We next consider an even simpler cloud structure without a
deep absorbing cloud.

\subsubsection{Fits of one-cloud models as a function of the \herat ratio}
\begin{figure*}\centering
\hspace{-0.3in}\includegraphics[width=3.25in]{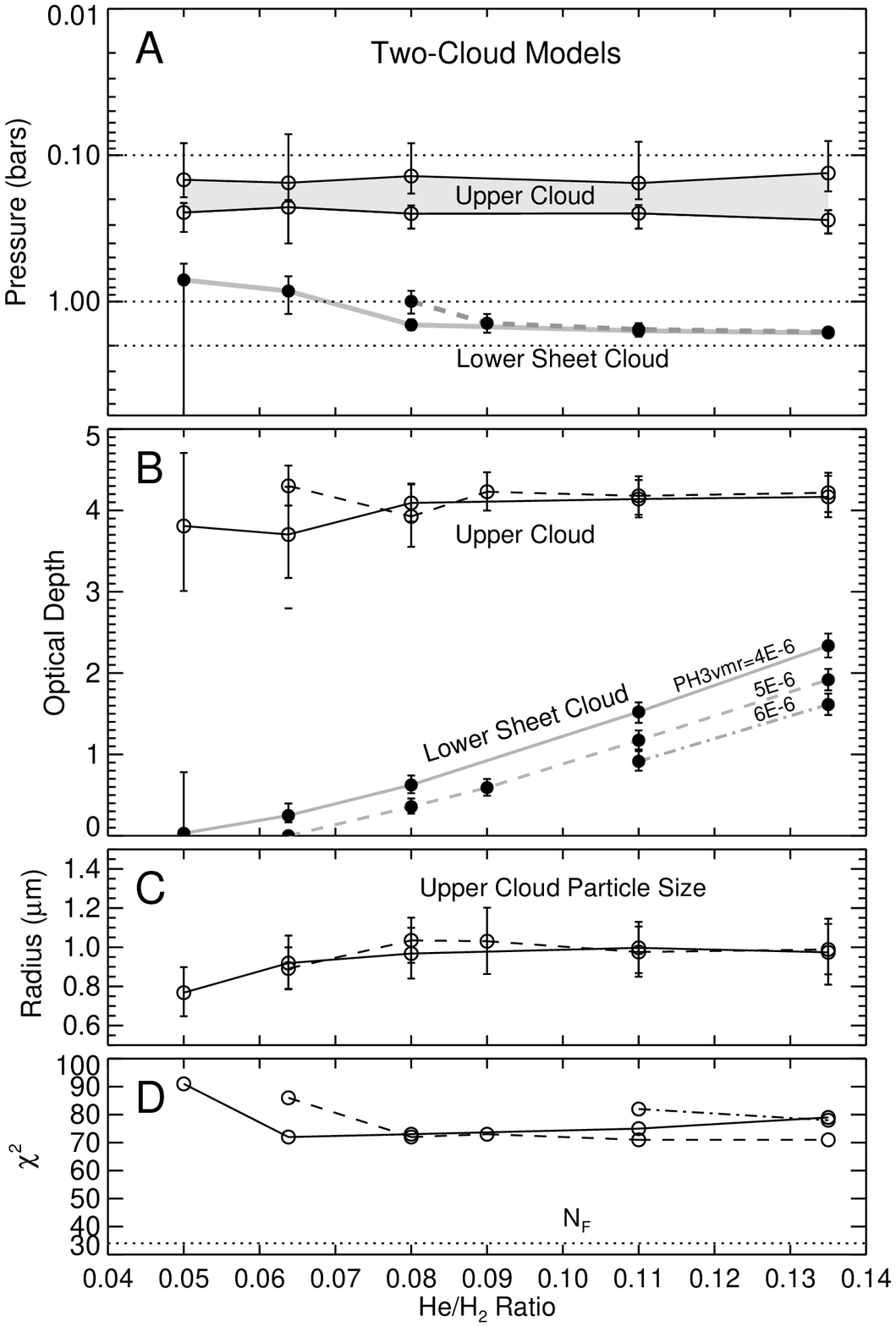}
\hspace{-0.1in}\includegraphics[width=3.25in]{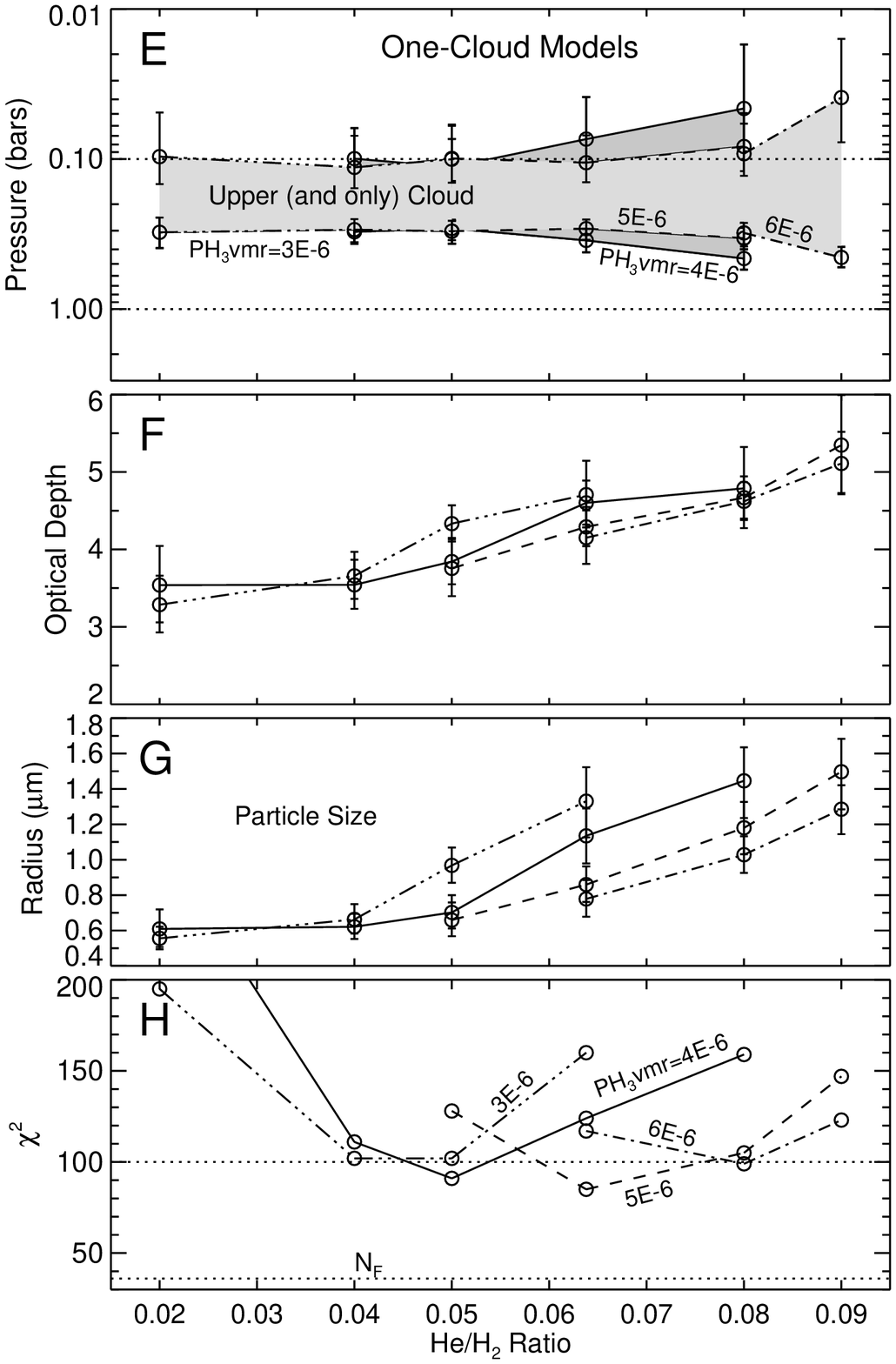}
\caption{Cloud structure inferred from one of the clearest regions
on Saturn as a function of the assumed He/H$_2$ ratio for two-cloud
models (A-D, left) and one-cloud models (E-H, right).  Note that the
\chisq values for the two-cloud model are nearly flat vs. \herat ratio and reach
a lower minimum (70) than the one-cloud model (86). The reduced \chisq values
(\chisq /N$_F$) are 2.09 and 2.39 respectively. \label{Fig:corefits}}
\end{figure*}

Next we consider solutions for which the deep sheet cloud is entirely
absent, solutions for which there are no aerosols present between the
upper cloud and the 5-bar region, the approximate location of the peak
in the thermal emission contribution function.  These fits are shown in
Fig.\ \ref{Fig:corefits}E-H.  In this case, the optical depth of the
upper cloud does vary with the \herat ratio, as does its particle
size.  The \chisq plot (H) shows that there is also a strong
preference for a \herat ratio $\approx$0.064 and a \pht
VMR$\approx$5\tenmsixx.  With \herat ratios 0.03 greater or 0.03
smaller than this value fit quality becomes dramatically worse no
matter what value of the \pht mixing ratio is chosen.  However, this
result is somewhat misleading because it is a consequence of how we
adjusted the \pht mixing ratio profile.  By varying the entire profile
by the same fraction as the deep mixing ratio we actually made it
difficult to fit the \pht absorption feature between 4.1 and 4.5
\mumx, which is a feature in reflected sunlight that is not much
affected by changing the \herat ratio, as evident from the lack of
variation shown in Fig.\ \ref{Fig:brightfit}.  Thus there is no
benefit to varying the \pht profile at low pressures (100 -- 500 mbar)
when changes need to be made at higher pressures to control thermal
emission. In the following section we discuss a different style of \pht profile
adjustment, in which only deeper mixing ratios are varied.

\section{Using the one-cloud model to constrain absorbing gas mixing ratios}\label{Sec:constraints}

If we are to explain the great longitudinal uniformity and high 5-\mum
brightness that developed in the wake of the Great Storm as a
dynamical clearing of cloud particles below the upper cloud, then the
single-cloud model fits tell us under what conditions that explanation
is possible. For our previously assumed style of \pht variation, and
without varying any other trace gases, the best matches to VIMS
spectra appeared to occur with a \herat ratio $\approx$
0.064$\pm$0.02.  However, the range of acceptable \herat ratios can be
expanded considerably if we adjust just the deep \pht profile and also
allow arbitrary adjustment of the arsine and ammonia mixing ratios.
The following two extreme cases provide instructive examples.

\subsection{Adjusting gas profiles to improve fits with \herat = 0.02.}

\begin{figure}\centering
\hspace{-0.3in}\includegraphics[width=3.5in]{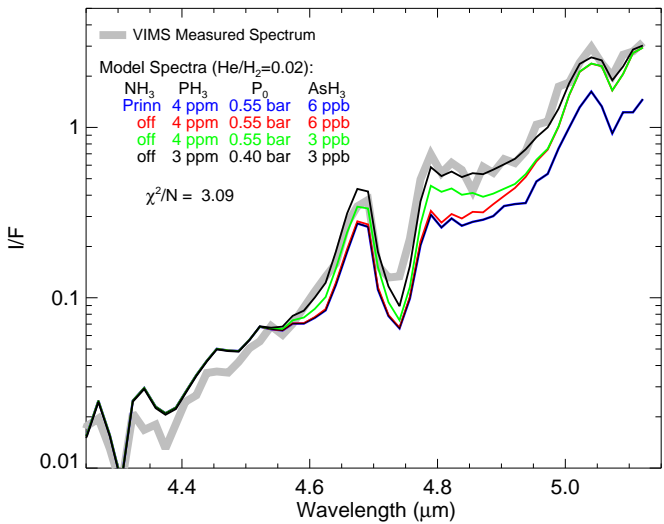}
\caption{Modifications of absorbing gas mixing ratio profiles to
  obtain optimum fits in the brightest wake region assuming a \herat
  ratio of 0.02 and no deep absorbing cloud. In this case greatly
  reduced absorption is needed to boost the model I/F to the observed
  level (gray curve). The best-fit vertical mixing ratio profile for
  \pht is shown as a dotted line in Fig.\ \ref{Fig:ph3profs}. The \nht
  profile for this case must be much less than the standard profile,
  but not constrained to a specific value. All the model profiles for
  \pht use $f$ = 0.5.
 \label{Fig:emit02}}
\end{figure}

The one-cloud model from
Fig.\ \ref{Fig:cartoon} was optimized for \herat= 0.064 and the
nominal trace gas profiles. If we use that cloud model to compute a spectrum for the
thermal structure appropriate to a \herat ratio of 0.02, we see from
blue curve in Fig.\ \ref{Fig:brightfit} that the resulting spectrum
remains a good fit in reflected sunlight, but falls a dramatic factor
of two short of the observed I/F in the thermal emission region.  This
discrepancy is shown in greater detail in Fig.\ \ref{Fig:emit02},
where the blue curve is for the same model shown by that color
in Fig.\ \ref{Fig:brightfit}.  The other curves in
Fig.\ \ref{Fig:emit02} show how that discrepancy can be fixed, first by
drastically reducing the \nht mixing ratio (producing the red curve),
then by also reducing just the deep \pht mixing ratio (producing
the green curve) and finally by dropping the arsine volume mixing ratio
to 3 ppb (producing the black curve).  
Because \nht is the dominant
absorber for $\lambda >$ 5.1 \mum (see Fig.\ \ref{Fig:pendepth}), the
\nht mixing ratio had to be reduced to negligible levels to boost
emission in that region. 
These modifications result in a \chisqx/N of 3.09, evaluated for the 38
spectral points beyond 4.5 \mumx.  The corresponding value for the
\herat = 0.064 spectrum is a significantly worse 4.55, even after some
fine tuning of the scale height ratio.

\begin{figure}\centering
\includegraphics[width=3.5in]{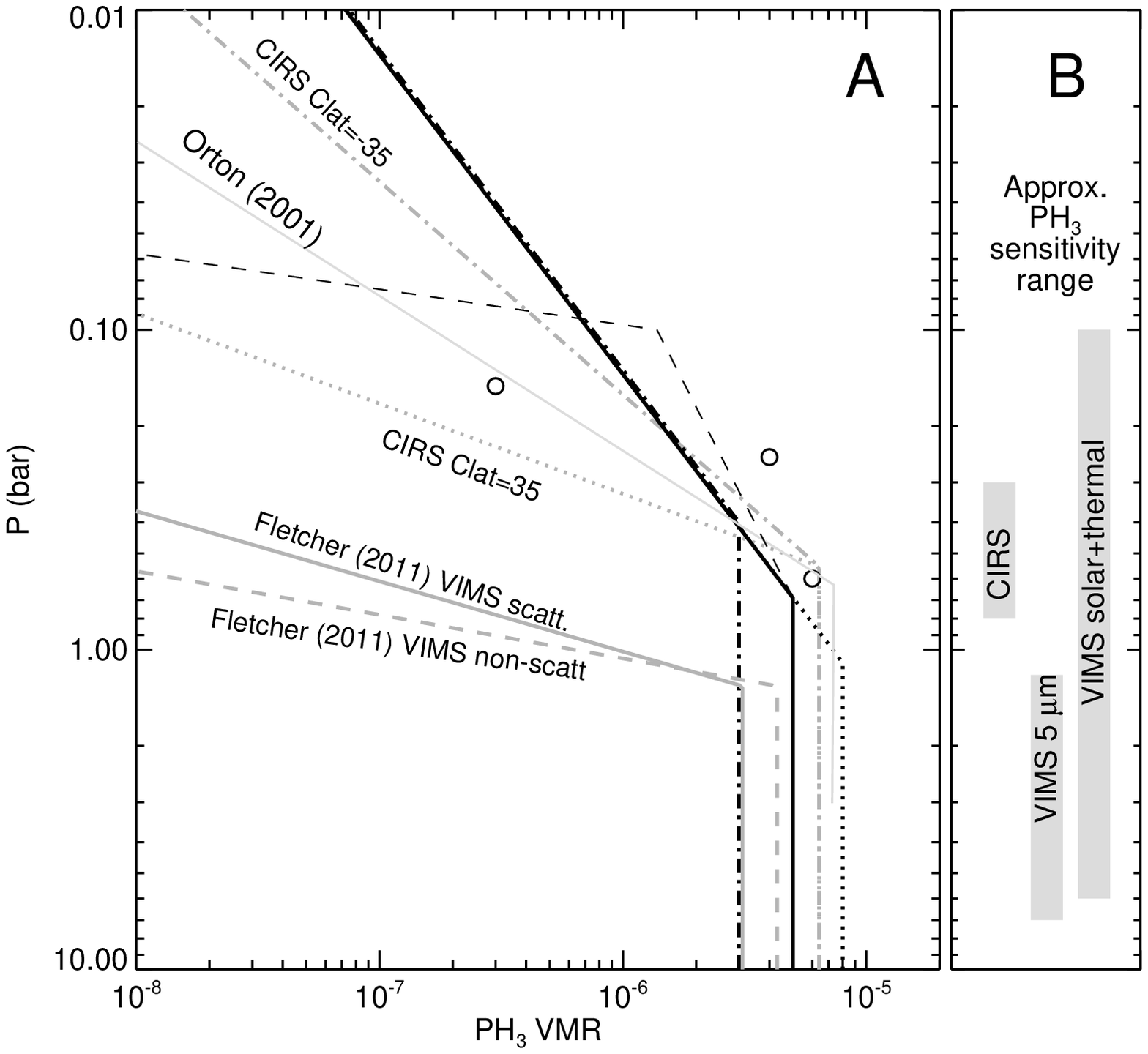}
\caption{A: Phosphine profiles we inferred from fitting the brightest
  ``cleared'' region of the wake assuming no deep absorbing cloud, are
  shown as dark lines, using solid for \herat = 0.064, dot-dash for
  \herat = 0.02, dotted for \herat = 0.135, and dashed for the
  three-slope profile yielding the same fit quality as the solid
  profile, but is more compatible with the sharper cutoff of the
  \cite{Lellouch2001} results (open circles). For comparison phosphine
  profiles from CIRS \citep{Fletcher2009ph3} and VIMS
  \citep{Fletcher2011vims} are also shown using lighter lines.  The
  \cite{Orton2001ph3} result (lightest gray line) is based on sub-mm
  groundbased observations and represents a disk average. The CIRS and
  \cite{Fletcher2011vims} results are based entirely on thermal
  emission spectra and assume a \herat ratio of 0.135.  B: Sensitivity
  ranges (to \phtx) for the CIRS and VIMS 5-\mum spectra are according
  to Fig. 6 of \cite{Fletcher2011vims}. The sensitivity range
  indicated for VIMS solar+thermal spectra is bounded by twice the
  unit optical depth range from our Fig.\ \ref{Fig:pendepth} for regions
  of the spectrum where phosphine is dominant, e.g. near 2.8 -- 3 \mum
  and 4.1 -- 5.1 \mumx. Note the order of magnitude disagreement
  between the CIRS and \cite{Fletcher2011vims} VIMS results in the 800
  mb region. In the same region our VIMS results are relatively
  consistent with CIRS results, considering the substantial
  variability that has been observed.
 \label{Fig:ph3profs}}
\end{figure}

Note that our above adjustment of the \pht profile was
designed to affect the thermal emission without affecting the
reflected solar model by keeping the mixing ratio the same
at lower pressures and preserving the falloff rate down to the
level at which it intersects the chosen deep mixing ratio.
To change the deep mixing ratio from  $\alpha_0$ to  $\alpha_x$ without
changing the VMR profile above the original break-point, we pick a new
pressure break-point $P_x = P_0 ( \alpha_{x}/\alpha_0)^{f/(1 - f)}$, 
where $f$ is the original scale height ratio.  The profile we selected
to optimize the model spectrum for \herat = 0.02 is shown by the
black dot-dash profile in Fig.\ \ref{Fig:ph3profs}, where the nominal
profile (for \herat= 0.064) is shown as a solid black line.

\subsection{Adjusting gas profiles to improve fits with \herat = 0.135.}

\begin{figure}\centering
\hspace{-0.3in}\includegraphics[width=3.5in]{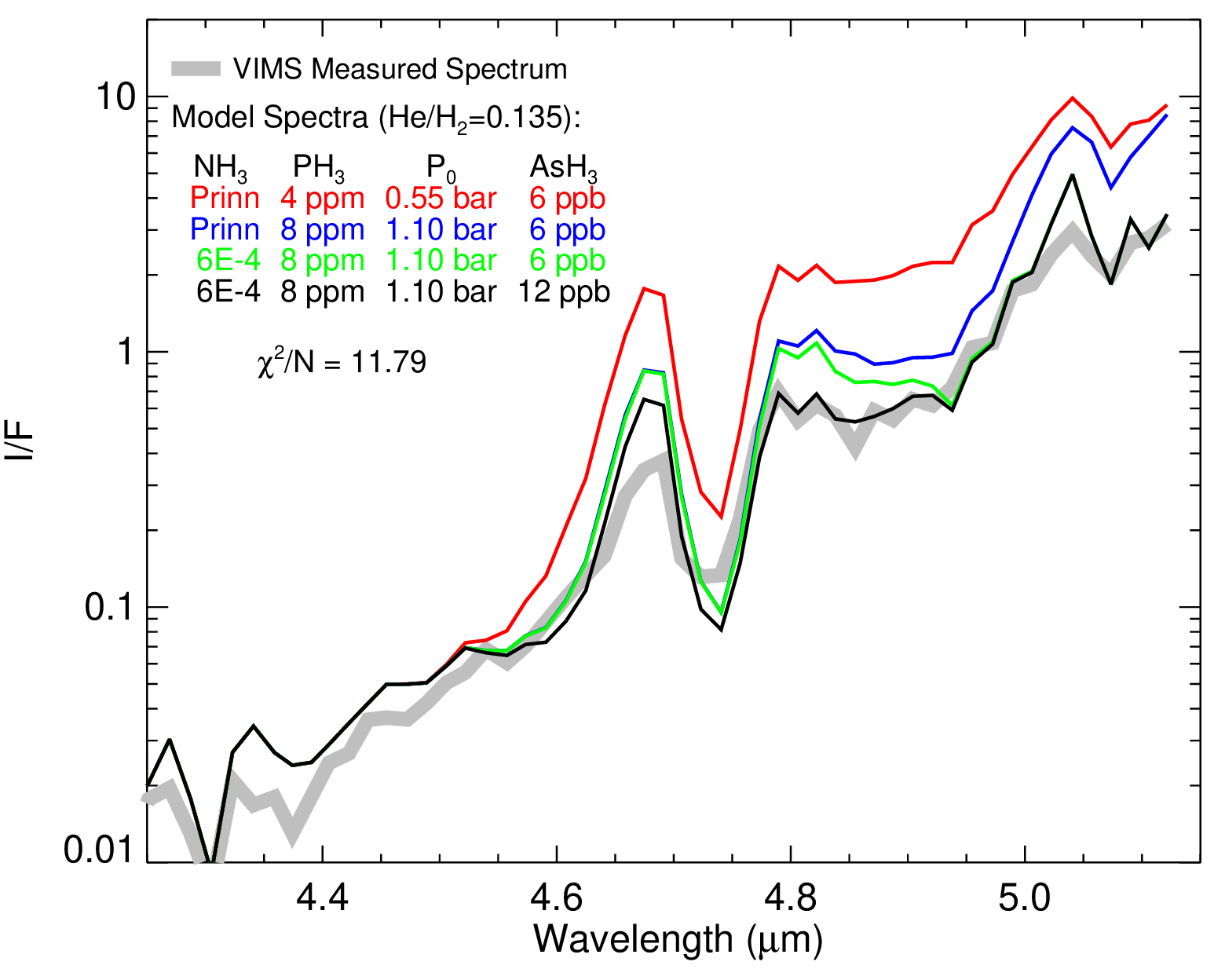}
\caption{Modifications of absorbing gas mixing ratio profiles to
  obtain optimum fits in the brightest wake region assuming a \herat
  ratio of 0.135 and no deep absorbing cloud. In this case greatly
  increased absorption is needed to reduce the model I/F to the
  observed level (gray curve). The best-fit vertical mixing ratio
  profiles for this case are shown as dot-dash lines in
  Fig.\ \ref{Fig:ph3profs} for \pht and in Fig.\ \ref{Fig:ammonia} for
  \nhtx. All model profiles for \pht use $f$ = 0.5.
 \label{Fig:emit135}}
\end{figure}

\begin{figure}\centering
\includegraphics[width=3.5in]{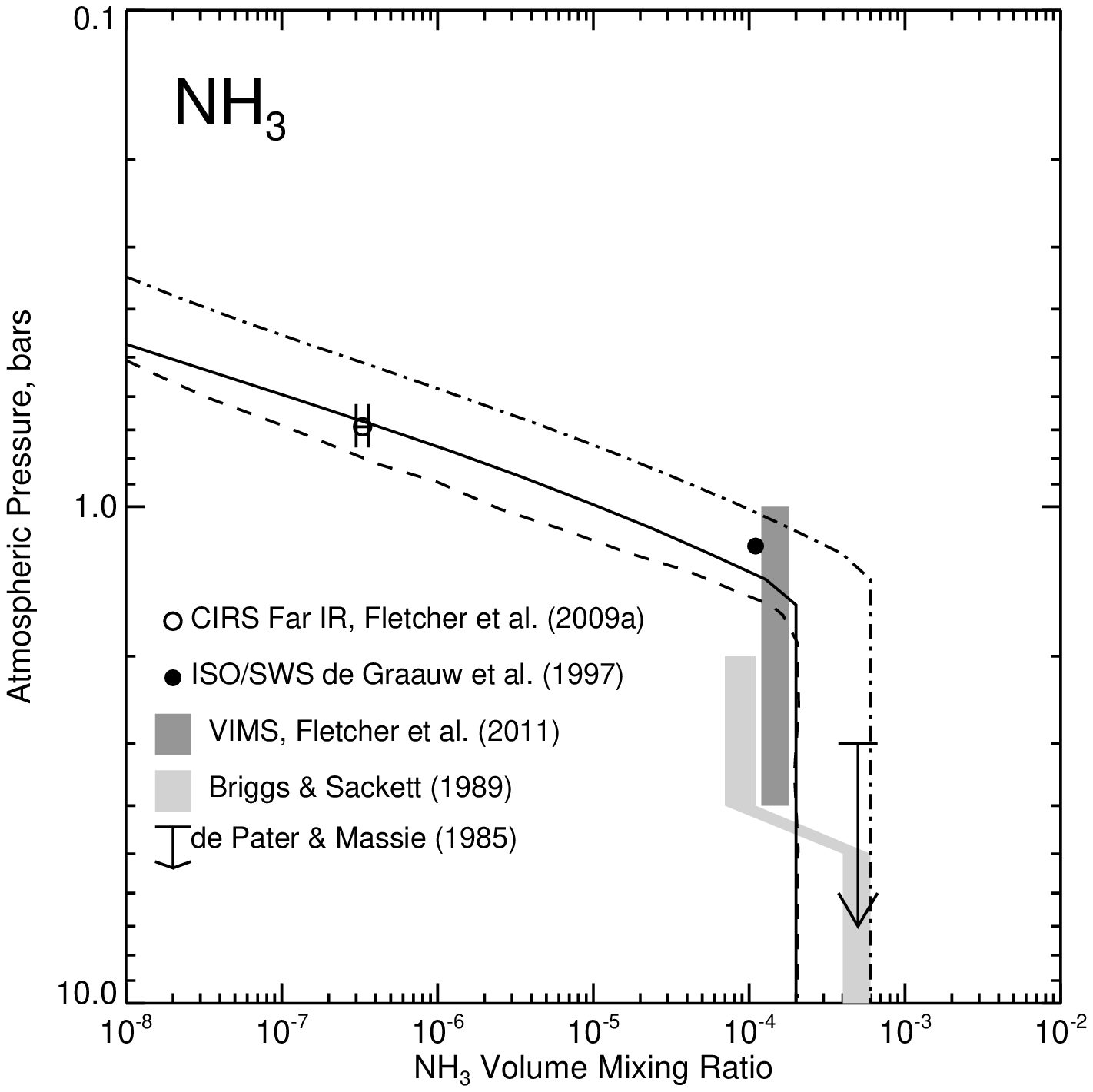}
\caption{Ammonia profiles we used to adjust emitted radiation to be
  consistent with the brightest clear wake spectrum are indicated by
  dashed (for \herat = 0.064) and dot-dash (for \herat = 0.135). The
  dashed curve is from \cite{Prinn1984} and the solid curve shows the
  saturated vapor profile with the same deep mixing ratio. For \herat
  = 0.02 no profile is plotted because we assumed ammonia was depleted
  by more than an order of magnitude, and thus negligible. Independent
  observations  are identified in the legend and discussed in the
  text.  \label{Fig:ammonia}}
\end{figure}

In this case, using the nominal one-cloud model to compute a spectrum
for a thermal structure appropriate to \herat = 0.135, we see from the
red curve in Fig.\ \ref{Fig:brightfit} that although the reflected
sunlight dominated part of the spectrum remains well fit, the model
spectrum exceeds observations by more than a factor of three in the
thermal emission region.  A more detailed view of this region is
displayed in Fig.\ \ref{Fig:emit135}, where the red curve is for the
same model represented by that color in
Fig.\ \ref{Fig:brightfit}. This figure also shows how the large
discrepancy can be greatly reduced by increasing the deep \pht VMR
from 4 ppm to 8 ppm and increasing the pressure break point from 550
mbar to 1.1 bars (producing the blue curve), by also increasing the
deep \nht mixing ratio to 6$\times 10^{-4}$ (producing the green
curve), and finally, by doubling the arsine VMR to 12 ppb (producing
the black curve).  The \pht profile we used here is displayed by the
black dotted curve in Fig.\ \ref{Fig:ph3profs}. The adjusted ammonia
profile is shown by the dot-dash curve in Fig.\ \ref{Fig:ammonia},
where the nominal \cite{Prinn1984} profile is shown by the dashed
curve.  The final model spectrum is a poorer fit than we were able to
achieve for low values of \heratx, and required an amount of \nht that
we will show exceeds other independent estimates in the critical 1 --
4 bar region.  Thus, this solution seems less plausible.

\subsection{Direct spectral comparison of fits at different \herat ratios.}

\begin{figure*}\centering
\hspace{-0.15in}
\includegraphics[width=2.75in]{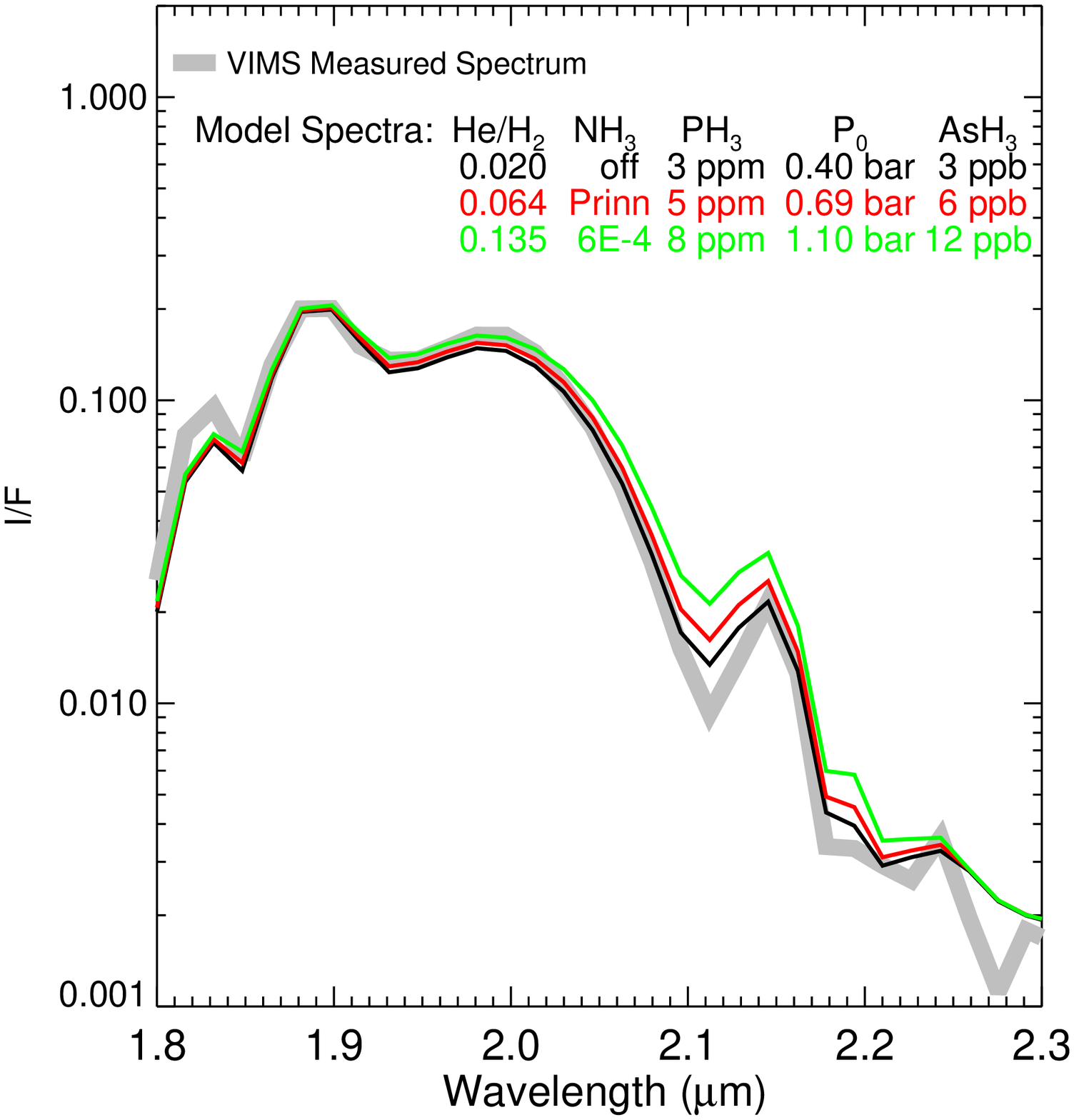}
\hspace{-0.1in}
\includegraphics[width=3.7in]{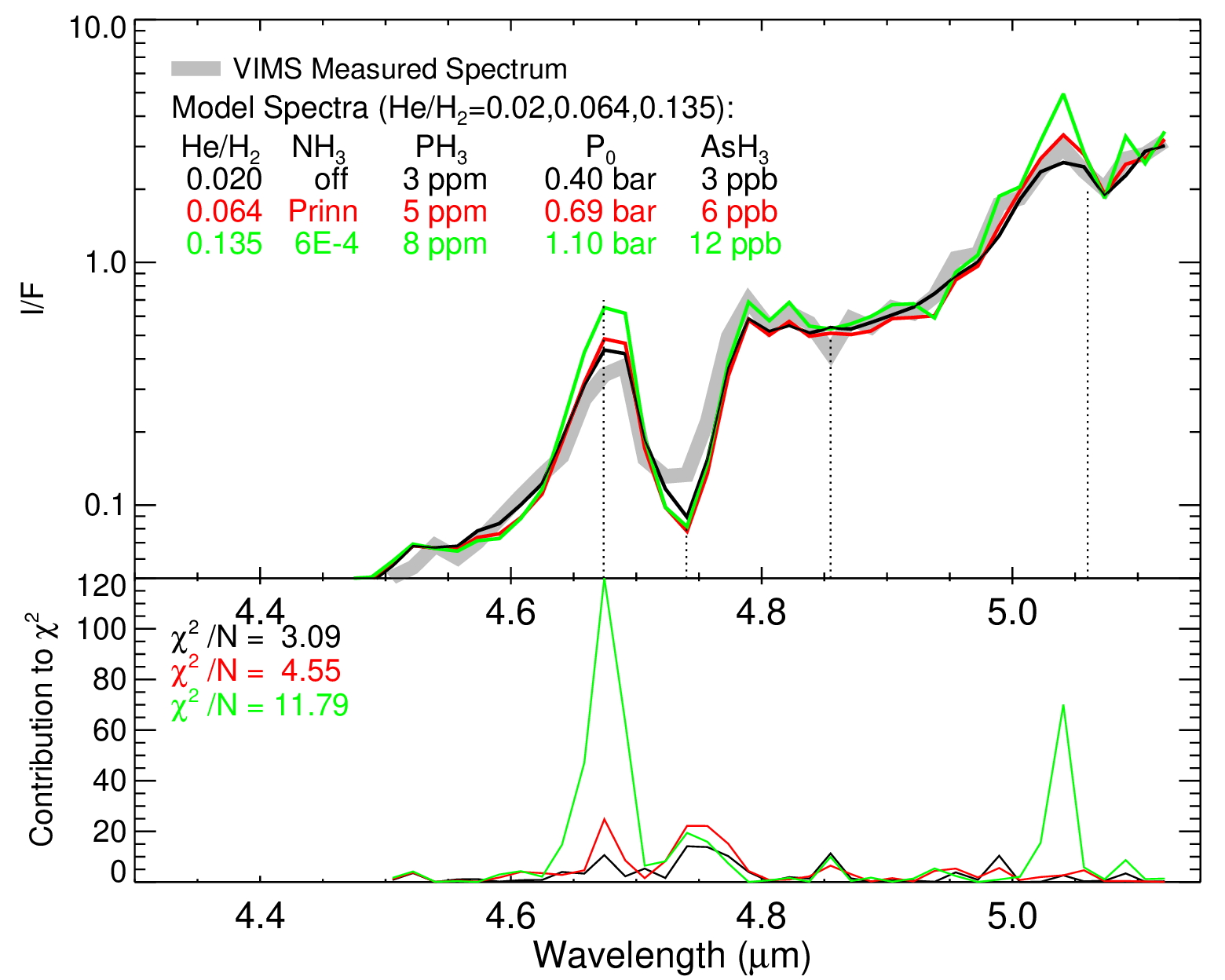}
\caption{Comparison of one-cloud best-fit models of the brightest wake
  thermal emission spectrum for three different \herat ratios, at both
  solar-dominated wavelengths (left panel) and thermal-dominated
  wavelengths (right panel).  The vertical mixing ratio profiles for
  \pht and \nht for these models are shown in
  Figs.\ \ref{Fig:ph3profs} and \ref{Fig:ammonia} respectively.  All
  models use $f$ = 0.5. Vertical dotted lines are plotted in the right
  panel to identify regions of spectral fitting problems discussed in
  the text.
 \label{Fig:emitmulti}}
\end{figure*}

Cloud structure and gas mixing ratio models were optimized to match
the thermal emission spectra without disturbing fit quality in
the reflected solar dominated part of the spectrum. However, these
actually did lead to small differences near 2.1 \mum as a result of
different degrees of hydrogen absorption that are visible in this
methane window region (see Fig.\ \ref{Fig:pendepth}).  Fig.
\ref{Fig:emitmulti} compares fits for three different \herat ratios
both in the 1.8-2.3 \mum region (left panel) and in the thermal
emission dominated region (right panel). Both comparisons seem to
favor the lower \herat ratios in overall fit quality.  However, some
of the small scale features in the measurements in the 4.8-5.15 \mum
region are smoothed over too much by the \herat = 0.02 fit, and better
matched by \herat = 0.064 fit.

We found no models that were able to provide a good fit at both 4.67
\mum and 4.74 \mumx.  Most models produce spectra that were too
high at the former and too low at the latter, and this wavelength region
contributed a substantial fraction of the $\chi^2$ for the thermal
fits (lower right panel of Fig.\ \ref{Fig:emitmulti}).  Nor was any
model able to reproduce the small dip near 4.85
\mumx. \cite{Fletcher2011vims} also noted similar problems at 4.67
\mum and 4.85 \mumx, as well as an under-fitting problem at 5.06
\mumx, which did not stand out in our modeling.  At shorter
wavelengths we also noted problems in fitting the 4.3-\mum \pht band,
for which models tended to be more asymmetric than the observations,
as can be seen in Fig.\ \ref{Fig:brightfit}. We also were unable to
closely reproduce the depth of the methane absorption feature at
2.58-\mumx, also visible in Fig.\ \ref{Fig:brightfit}. Whether any of
these problems might be resolved by improved line data remains to be
determined.

\section{Using independent gas constraints to constrain \herat ratios.}

We have shown that the lack of a deep absorbing cloud in the putative
clear regions increases thermal emission at high \herat ratios and
decreases thermal emission at low \herat ratios.  We have also shown
that these changes can be largely compensated for by adjusting the
vertical profiles of \pht and \ashtx.  Here we consider other
constraints on these gas profiles that might be inconsistent with such
adjustments, and thus provide limits on the range of acceptable
\herat ratios.

\subsection{Limits to phosphine adjustments}

\subsubsection{CIRS-VIMS comparisons}

In Fig.\ \ref{Fig:ph3profs}A, the \pht profiles we derived for the  clearest
region of the wake are compared with CIRS results of
\cite{Fletcher2009ph3} and night-side VIMS results of
\cite{Fletcher2011vims}. As indicated by
gray bars in the figure (panel B), CIRS results are sensitive to \pht in the 400
-- 800 mbar range, while VIMS night-side spectra are sensitive to
\pht in the 1.2 -- 7 bar range (roughly), according to Fig. 6 of
\cite{Fletcher2011vims}.  While CIRS results include a substantial
latitudinal variation, here we show only example results from 35\deg N
and 35\deg S (planetocentric).  The night-side VIMS results of
\cite{Fletcher2011vims} are shown only for 35\deg N and for two
different retrieval models, one with a non-scattering gray absorbing
cloud and a second that includes a scattering cloud.  Both models have
smaller deep VMR values and smaller scale heights than other results.

Although these \cite{Fletcher2011vims} night-side retrievals of \pht
show considerable variability depending on what type of cloud model is
assumed, the large discrepancy between the night-side results and the
CIRS results in the 400 -- 800 mb range is a robust characteristic.
This is surprising, especially given that our inferred \pht profile is
in much better agreement with CIRS results in the range where CIRS is
sensitive to \phtx. Most of these profiles are characterized by a
constant mixing ratio for $P>P_b$ and a decline with altitude with a
scale height that has a constant ratio to the pressure scale height,
following Eq.\ref{Eq:prof1}. The exception is the three slope profile
shown by the black dashed curve in Fig.\ \ref{Fig:ph3profs}. In that
case there are two pressure break points $P_0$ and $P_1$ and two scale
heights $f$ and $f_1$, and the upper region of the profile satisfies
Eq.\ \ref{Eq:prof2}, while Eq.\ \ref{Eq:prof1} applies for
$P_0>P>P_1$.

\subsubsection{Constraints from the 4.1 -- 4.6 \mum band}

\begin{figure}\centering
\includegraphics[width=3.5in]{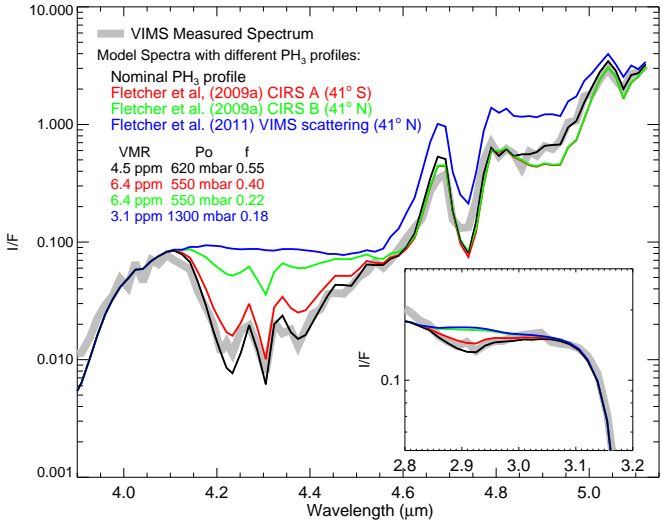}
\caption{Model spectra for different \pht profiles shown in
  Fig.\ \ref{Fig:ph3profs}.  All profiles assume a uniform deep VMR
  for $P \ge P_0$ and an exponential decline for $P \le P_0$ with a
  \pht to pressure scale height ratio of $f$, with specific parameter
  values shown in the inset table. Except for the
  \cite{Fletcher2011vims} VIMS-based profile, all the \pht profiles yield similar
  thermal emission spectra.  But very large differences are seen at
  reflected solar wavelengths, which are sensitive to \pht mixing
  ratios at lower pressures as well as higher pressures (inset). Note
  that the observed 4.3-\mum \pht band is measured to be more
  symmetric about 4.3 \mum than our models indicate.
 \label{Fig:ph3specs}}
\end{figure}

The large discrepancy between CIRS and VIMS \pht profiles is
especially easy to detect in the reflected solar spectrum, as
illustrated in Fig.\ \ref{Fig:ph3specs}. Here we see large differences
in spectral shape in the 4.1 -- 4.5 \mum region, with the best match
to observed spectra occurring with our nominal profile and the CIRS B
profile (which has a larger scale height fraction derived from 35\degx
S).  The 4.1 -- 4.6 \mum wavelength range is where reflected sunlight
is especially sensitive to the \pht mixing ratio in the 100 -- 500
mbar pressure range.  The night-side VIMS profile of
\cite{Fletcher2011vims} produces essentially no detectable absorption
feature in this spectral region. It also produces excessive I/F values
at thermal emission wavelengths because of its relatively low deep VMR
(for its assumed \herat ratio of 0.135).  The spectral matches to the
much smaller absorption feature near 2.9 \mum are consistent with what
is seen at longer reflected solar wavelengths.  These results make a
strong case for using both solar and thermal spectral regions in
combination to constrain the vertical profile of phosphine. The 4.1 --
4.5 \mum absorption band is evidently an important one for
constraining \pht on Saturn, but has seen little use so far.

\subsubsection{Constraints from ISO and groundbased observations}

 Disk-averaged values have been derived by a number of observers based
 on high spectral resolution observations that can resolve more
 distinctive spectral features of \pht in the thermal emission range
 of Saturn's spectrum and are thus probably less sensitive to
 different assumptions of \herat ratios and less sensitive to assumed
 cloud properties. \cite{Lellouch2001} used ISO/SWS observations of
 the 8.1 -- 11.3 \mum spectrum to infer \pht mixing ratios in the 100
 -- 600 mbar range, with a value of 6 ppm up to 600 mbar, 4 ppm at 250
 mb, decreasing to 0.3 ppm at 150 mbar. These are plotted as open
 circles in Fig.\ \ref{Fig:ph3profs}.  Also shown there using a gray
 solid line is the profile inferred by \cite{Orton2001ph3} from sub-mm
 thermal emission observations.  The falloff of \pht VMR with altitude
 is an expected result of photochemical destruction. Although the falloff rate
 is not well defined, it is probably sharper than most of these
 profiles indicate.

\begin{figure}\centering
\includegraphics[width=3.5in]{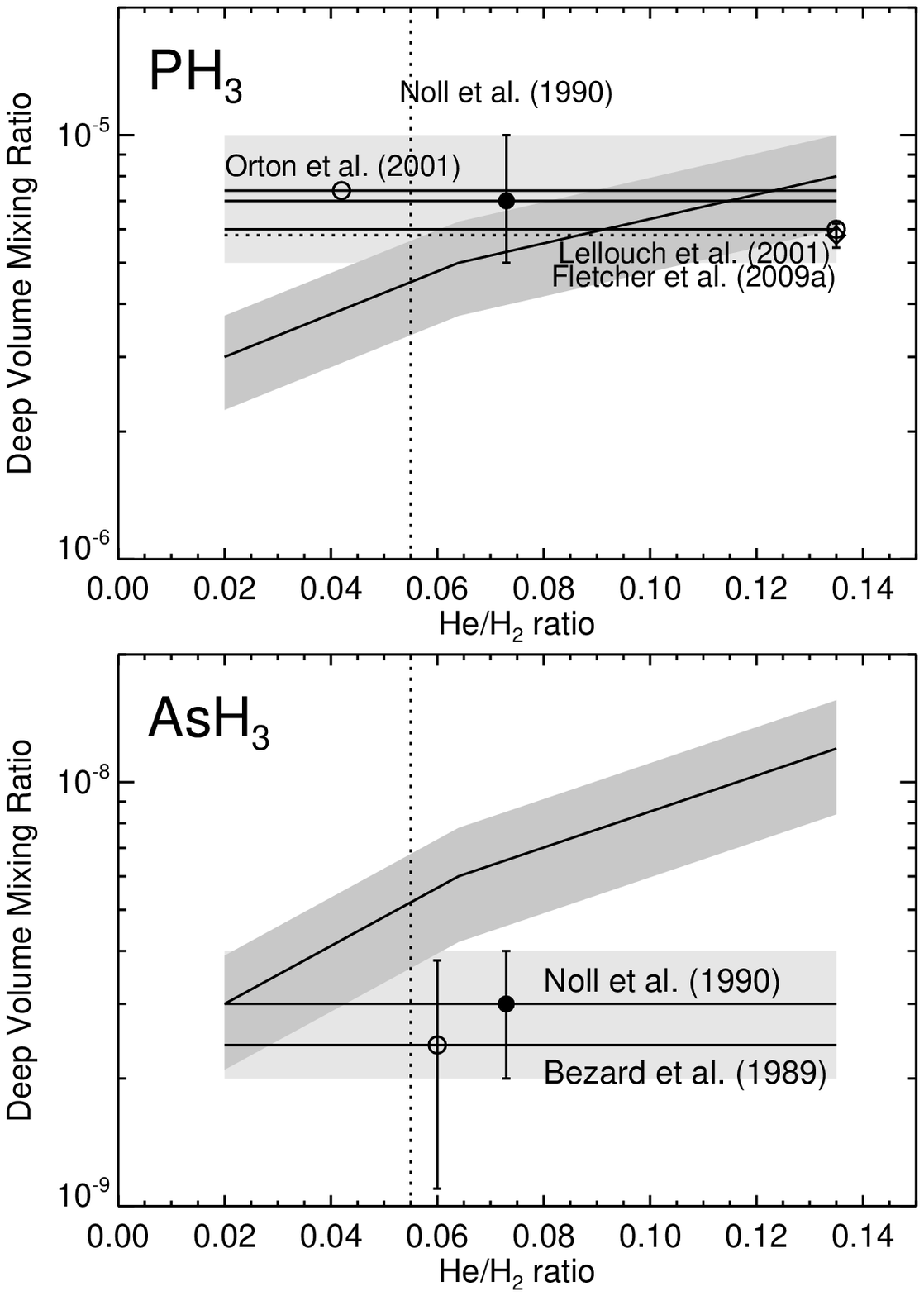}
\caption{Deep mixing ratios of \pht and \asht as a function of the
  \herat ratio.  Solid curves with darker gray uncertainty bands are
  from our analysis of the brightest wake spectrum assuming that there
  is no deep absorbing cloud. Independent observations of \pht and
  \asht provide possible limits on the \herat ratios. The independent
  values for \pht and \asht are plotted at the \herat ratio that was
  used in each analysis.  It is not clear how much a given analysis
  would change if a different \herat ratio had been assumed. The vertical
  dotted lines are plotted at a \herat ratio of 0.055, which is a
  possible compromise that crudely satisfies both constraints. See
the main  text for further discussion.
 \label{Fig:vmr_vs_herat}}
\end{figure}

Fig.\ \ref{Fig:vmr_vs_herat} displays deep \pht VMR values as a
function of the \herat ratio assumed in our analysis of the VIMS emission
spectrum from the brightest region of the wake. Also shown are
independent determinations of the ratio from high-resolution 5-\mum
spectra by \cite{Noll1990}, from ISO-SWS spectra by
\cite{Lellouch2001}, and from sub-mm spectra by
\cite{Orton2000ph3err,Orton2001ph3}.  \cite{Orton2000ph3err} note that
a 25\% reduction in \pht (i.e. from 7.4 ppm to 5.6 ppm) would be
produced by a 2\deg decrease in the assumed temperature at all
pressures, which they noted would provide a slightly better fit for
the 3-2 line.  It would also provide better agreement with
\cite{Lellouch2001} and \cite{Fletcher2009ph3}. No specific error
estimates were provided for the Lellouch et al. and Orton et
al. results.  We also show the \cite{Noll1990} result which is a
disk-averaged result based on high-resolution groundbased observations
of the 5-\mum spectrum made in 1994.  Also shown is CIRS-based result
of \cite{Fletcher2009ph3}, which is an average of northern hemisphere
latitudes from 10\degx N to 70\degx N.  The latter two have error
estimates, although there is considerable uncertainty as to how these
results might differ if the authors had chosen different \herat ratios
in conducting their analyses.  If taken at face value, these results
suggest that the deep \pht mixing ratio is not likely to be as low as
seems to be required to match our VIMS results for \herat ratios below
$\sim$0.045.  The results are quite different for arsine.

\subsection{Limits to arsine adjustments}

Arsine measurements are plotted in the lower panel of
Fig.\ \ref{Fig:vmr_vs_herat}.  \cite{Bezard1989AsH3} derived \asht
mixing ratios of 2.4$^{+1.4}_{-1.2}$ ppb for the thermal component and
0.39$^{+0.21}_{-0.13}$ ppb for the reflected solar component. The
latter is probably representative of the effective value in the 200 --
400 mbar range where they inferred a haze layer, while the former
applies to the deep mixing ratio. Because we are using spectra with
exceptionally high thermal emission values, the reflected solar
contribution at thermal emission wavelengths is not very important. In
addition, even at its absorption peak the two-way $\tau$ = 1 level for
\asht is deeper than 1 bar. Thus it is only the deep mixing ratio that
is of interest here.  \cite{Noll1989AsH3} inferred a value of
1.8$^{+1.8}_{-0.8}$ ppb.  These are more compatible with the values we
derived from the lower values of the \herat ratio.  As shown in
Fig.\ \ref{Fig:vmr_vs_herat}, the independent constraints on arsine
suggest that the \herat ratio should be less than 0.06, while
constraints suggested by \pht measurements suggest values higher than
0.045.  The compromise value of 0.055 is plotted as the vertical
dotted line in Fig.\ \ref{Fig:vmr_vs_herat}.

\subsection{Limits to ammonia adjustments}

Selected independent measurements of Saturn's ammonia mixing ratio
profile are shown in Fig.\ \ref{Fig:ammonia}. Microwave observations
provide sensitivity to the deep mixing ratio, and both
\cite{DePater1985} and \cite{Briggs1989} are in good agreement on a
value of 400 -- 600 ppm. \cite{Briggs1989} also note a decline in the
\nht VMR at pressures less than 5 bars and a value of 70 -- 110 ppm at
2 bars.  This provides the basis for our sketched profile in
Fig.\ \ref{Fig:ammonia}.  The 70 -- 110 ppm estimate covers the
pressure range most relevant for modeling Saturn's 5-\mum spectra. The
CIRS-based results of \cite{Fletcher2011vims}, shown as the dark gray
bar in Fig.\ \ref{Fig:ammonia}, are closer to the profiles we used for
the \herat ratio of 0.064, which is from \citep{Prinn1984} and shown
as the dashed curve. The nearby solid curve has the same deep mixing
ratio, but follows the saturated vapor profile above the saturation
level.  The dot-dash curve traces the profile we used to suppress
thermal emission in the model for \herat = 0.135.  This is well above
the global average values measured in the 2 -- 3 bar range, and is
perhaps the strongest argument against a \herat ratio much greater
than 0.06, at least under the assumption that the clearest wake region
does not contain a deep absorbing cloud.  In a region of unusually
high 5.1-\mum brightness it would indeed be implausible to have much
higher than normal levels of ammonia vapor. The profile we used to
raise the I/F at 5.1 \mum in the case of a \herat ratio of 0.02 is not
plotted here because ammonia had to be reduced to negligible levels in
that case. An upper bound for that case remains to be determined.  It
is not clear what an implausible level would be for ammonia in an
unusually clear region.  Downwelling motions of gas from above (lower
pressure than) the 1 bar level could certainly depress the ammonia
mixing ratio at deeper levels.  An example of such an effect was
observed by the Galileo probe that entered an unusually clear region
in Jupiter's atmosphere at the edge of a 5-\mum hot spot, where mixing
ratios of several condensable gases were depressed by roughly an order
of magnitude \citep{Niemann1998}, presumably by downwelling motions.
Thus, a highly depressed ammonia abundance in what appear to be a
Saturnian ``hot band'' is quite plausible.  An order of magnitude
depletion could be obtained by mixing gas downward from the 1 bar
level or above (lower pressure).  Also relevant are 2.2-cm emission
measurements from May 2011 \citep{Janssen2013, Laraia2013}, which
showed that the wake region was becoming "dried out" with respect to
ammonia vapor, supporting the conclusion that a depletion of \nht
clouds might be occurring, contradicting the idea that the region
might have greater than average amounts of ammonia.  The main message
from consideration of ammonia constraints is that low values of the
\herat ratio are more plausible than high values.

\section{Summary and conclusions}

The remarkably uniform and 5-\mum bright wake of Saturn's Great Storm of 2010 -- 2011
was investigated with the help of VIMS spectral images, using both reflected
sunlight portions of the spectrum and that part dominated by thermal emission ($\lambda > 4.5$ \mumx)
to constrain vertical cloud structure.  Our conclusions from this analysis
can be summarized as follows.

\begin{enumerate}

\item The wake region began with widespread appearance of absorption near 3 \mumx, was
generally very dark at thermal emission wavelengths (near 5 \mumx), with exceptions
of local regions that were brighter at 5 \mum than even before the beginning of the
storm.  The regions brightest near 5 \mum were near the anticyclone that developed
along with the convective storm feature.

\item As time progressed the regions of high 5-\mum brightness expanded longitudinally
beginning near the middle of the wake region,
eventually expanding latitudinally to cover the entire band from 29\degx N to 39\degx N,
and all longitudes.  In addition, by December 2012, the band became remarkably uniform,
with an RMS deviation over longitude of only 2\% in I/F at 5.12 \mum in the middle of the band,
and had similar latitudinal uniformity from 30\degx N to 39\degx N over most longitudes.

\item Before the storm began, VIMS spectra at the storm latitudes could be well fit
using model structure with mainly just two cloud layers. In the 190 -- 570 mbar range 
we inferred an upper layer of conservative particles of unknown composition 
and about 5 -- 6 optical depths, which scatter as spheres
with refractive index of n = 1.4+0i and particle radii slightly less than 1 \mumx.
A second, deeper layer of partially absorbing cloud particles was needed to limit thermal
emission. Assuming an arbitrary single-scattering albedo of 0.95, and a physically
thin sheet cloud structure, we found that it needed to be near 1.5 bars and have
an optical depth near 3.

\item Applying our two-cloud model to the clear regions of the wake over
  time we found that the upper layer optical depth dropped by almost a
  factor of two initially, then slowly grew over time. The more
  significant effect was on the lower cloud that dropped its optical
  depth initially by a factor of four, reaching a factor of five
  decrease in December 2012, reaching a minimum optical depth of 0.57.

\item While the above results indicate that the lower cloud never
  completely disappeared, the presence or absence of that cloud
  depends critically on the assumed value of the \herat ratio and on
  what is assumed for mixing ratios of \nhtx, \pht and \ashtx.  For
  high values of the \herat ratio, either a significant cloud is
  required in the cleared regions or a significant increase in
  absorbing gas mixing ratios is required.  For values less than
  0.064, the lower cloud is not required, but for very low \herat
  ratios the atmosphere becomes so cold that the observed 5-\mum
  emission cannot be reached without significant reductions in gas
  absorptions.

\item When we fit a one-cloud (upper cloud) model to the spectra for a
  range of \herat ratios, allowing the \pht profile to vary by a scale
  factor, we found that the best fit was at a \herat ratio near 0.064
  and a \pht deep mixing ratio of 5 ppm.  But, with more realistic
  adjustments of \pht mixing ratios at just the higher pressures, and
  also including adjustments of \nht and \ashtx, we were able to
  expand the range of viable \herat ratios.

\item To limit the range of plausible \herat ratios, we compared our
  adjusted gas profiles to independent measurements of those profiles,
  finding that \pht observations suggest that the \herat ratio should
  be greater than 0.05, while \asht observations suggest that the
  \herat ratio should be less than 0.06. A compromise value is
  0.055. Ammonia limits suggest that the \herat ratio should be less
  than 0.07 and greater than 0.02.

\item Another factor is how well the various options fit the observed
spectra.  The best overall fit in the 4.5 -- 5.15 \mum region was
  obtained with \herat = 0.02, although that fit suppresses small
  spectral features below the level observed, while the fit with
  \herat = 0.135 exaggerates those features and also produces a
  much worse overall fit, suggesting that a \herat $\sim$0.064 is a
  better choice. The best fits in the 2.1-\mum region are obtained with \herat =
  0.02 -- 0.064. In the net, fit quality favors \herat values in the 0.02 -- 0.064
range.

\item If the broad clearing of the wake region is to be explained as a complete
disappearance of lower cloud particles, and considering fit quality and all the
independent constraints on absorbing gases, it appears that the \herat ratio 
would need to be in the range of 0.055$_{-0.15}^{+0.10}$. 

\item We also showed that including reflected sunlight in the 4.1 -- 4.5 \mum \pht
  absorption band enabled VIMS observations to provide strong
  constraints on the \pht mixing ratio in the 100 -- 1000 mbar region. Our
  results in this region are roughly consistent with CIRS-based
  results of \cite{Fletcher2009ph3}, but disagree strongly with
  VIMS-based results of \cite{Fletcher2011vims}, by orders of
  magnitude in the 500 mbar region, where their deep pressure
  break-point and small scale height lead to much less \phtx.  Our
  scale height ratio to the pressure scale height, valid for the
  entire range of \herat values we considered was in the range of
  0.5 -- 0.6, compared to values $\sim$0.2 for \cite{Fletcher2011vims}.
  To allow for a sharper photochemical cutoff above 100 mbar, we
needed to increase the scale height even further to compensate
for lost absorption above that level, leading to a profile
similar to that inferred by \cite{Lellouch2001}, except for a
somewhat lower upper break-point pressure.

\item We identified spectral regions where persistent discrepancies between
model and observed spectra suggest the possibility of missing or
erroneous information in the trace gas line data. As indicated in Fig.\ \ref{Fig:emitmulti}, the
depth of the 4.74-\mum absorption feature in models always exceeds the
measured depth, and there is a small spectral absorption feature near 4.85 \mum 
in the measured spectrum that is completely absent from model spectra.
Also, as indicated in Fig.\ \ref{Fig:ph3specs}, the large 4.3-\mum absorption feature
is more asymmetric in models than in VIMS measurements, with models producing
less absorption on the long wavelength side of the absorption maximum.

\end{enumerate}

It remains to be seen how long it will take for the wake region to
return to the same state it had before the Great Storm began, or to
return to a state of reduced 5-\mum emission.  The
wake remained relatively uniform and bright at 5 \mum into 2015, although
some dimming seems to be underway \citep{Momary2015}.  Analysis of
its long term evolution beyond 2012 is left for future work.

\section*{Acknowledgments.} \addcontentsline{toc}{section}{Acknowledgments}

Support for this work was provided by NASA through its Cassini Data
Analysis and Participating Scientists (CDAPS) Program under grant
NNX15AL10G.  We thank Don Banfield and an anonymous reviewer for their
attention to detail and for providing many useful suggestions that
improved the manuscript.



\end{document}